\documentclass[final,5p, times, twocolumn]{elsarticle}
\usepackage[hidelinks,breaklinks=true]{hyperref} 
\hypersetup{colorlinks=true}
\usepackage{graphicx}
\usepackage[geometry]{ifsym}
\usepackage{amsmath} % align
\biboptions{sort&compress}

\journal{Journal of Nuclear Materials}

\begin{document}

\newcommand{\rmi}{\mathrm{i}}

%\preprint{}

\title{\textit{Ab initio} inspection of thermophysical experiments for zirconium near melting} %Title of paper



\author[jiht,mipt]{M. A. Paramonov}
\ead{mikhail-paramon@mail.ru}

\author[jiht,mipt]{D. V. Minakov\corref{cor1}}
\ead{minakovd@ihed.ras.ru}

\author[jiht,mipt]{V. B. Fokin}%
\ead{vladimir.fokin@phystech.edu}

\author[jiht,mipt]{D. V. Knyazev}%
%\ead{}

\author[jiht,mipt]{G. S. Demyanov}%
%\ead{}

\author[jiht,mipt]{P. R. Levashov}%
\ead{pasha@ihed.ras.ru}

\cortext[cor1]{Corresponding author}

\address[jiht]{Joint Institute for High Temperatures RAS, Izhorskaya 13 Bldg 2, Moscow 125412, Russia}%
\address[mipt]{Moscow Institute of Physics and Technology, 9 Institutskiy per., Dolgoprudny, Moscow Region, 141700, Russia}%

%\author{M.A. Paramonov}
%\affiliation{Joint Institute for High Temperatures RAS, Izhorskaya 13 Bldg 2, Moscow 125412, Russia}
%\affiliation{Moscow Institute of Physics and Technology, 9 Institutskiy per., Dolgoprudny, Moscow Region, 141700, Russia}
%\author{D. V. Minakov}
%\affiliation{Joint Institute for High Temperatures RAS, Izhorskaya 13 Bldg 2, Moscow 125412, Russia}
%\affiliation{Moscow Institute of Physics and Technology, 9 Institutskiy per., Dolgoprudny, Moscow Region, 141700, Russia}
%\author{P. R. Levashov}
%\affiliation{Joint Institute for High Temperatures RAS, Izhorskaya 13 Bldg 2, Moscow 125412, Russia}
%\affiliation{Tomsk State University, 36 Lenin Prospekt, Tomsk 634050, Russia}

%\thanks{}
%\affiliation{Joint Institute for High Temperatures RAS, Izhorskaya 13 Bldg 2, Moscow 125412, Russia}
%\affiliation{}


\date{\today}

\begin{abstract}

We present quantum molecular dynamics calculations of thermophysical properties of solid and liquid zirconium in the vicinity of melting. An overview of available experimental data is also presented. We focus on the analysis of thermal expansion, molar enthalpy, resistivity and normal spectral emissivity of solid and liquid Zr. Possible reasons of discrepancies between the first--principle simulations and experiments are discussed. Our calculations reveal a significant volume change on melting in agreement with electrostatic levitation experiments. Meanwhile, we confirm a low value of enthalpy of fusion obtained in some pulse-heating experiments. Electrical resistivity of solid and liquid Zr is systematically underestimated in our simulations, however the slope of resistivity temperature dependencies agrees with experiment. Our calculations predict almost constant normal spectral emissivity in liquid Zr. 

\end{abstract}

%\pacs{}% insert suggested PACS numbers in braces on next line

\maketitle %\maketitle must follow title, authors, abstract and \pacs

\section{Introduction}

Knowledge of high-temperature thermophysical properties of nuclear energy materials is essential for analyzing safety of nuclear reactors currently in operation or planned for operation in the future. Equations of state for reactor materials are crucial for modeling the behavior of nuclear power plants under critical impacts, predicting conditions that can lead to severe industrial accidents and for analysis of their consequences.
In this regard, accurate knowledge of the conditions at which a phase transition occurs in the system is of particular importance. This may be a change in the type of a crystal lattice, melting or evaporation.

Zirconium is some of the main structural materials of nuclear power plants, and is specifically used for the fuel elements of boiling water reactors, owing to its low neutron absorption cross section and resistance to corrosion.
%Gheribi 2009
%Simone Anzellini 2020
%Arblaster 2013
It has a hexagonal close-packed (hcp) structure ($\alpha$-phase) under normal conditions which transforms to a hexagonal structure ($\omega$-phase) under pressure \cite{tonkov2018phase, sikka1982omega}. Upon further compression at room temperature, Zr takes on the bcc structure ($\beta$-phase) \cite{dewaele2015high, tonkov2018phase, Anzellini:PRB:2020}. 
The hcp to bcc transition in Zr also occurs at heating to temperatures higher than about 1135~K~\cite{parisiades2019melting}.
%This is a first order martensitic transition, which proceeds via the Nishayama–-Wassermann mechanism (in paper Gheribi). 
%At the $\alpha$--$\beta$ transition, 
It is interesting, that anharmonicity in lattice vibrations stabilizes $\beta$-Zr at high temperatures~\cite{hellman2011lattice}. This circumstance significantly complicates a theoretical description of the $\beta$-phase at high temperatures, since harmonic models commonly used for crystalline materials become inapplicable.

Zirconium melts at about $T_m = 2128$~K at atmospheric pressure. High melting temperature and the risk of contamination of liquid Zr make it difficult to measure its thermophysical properties at $T > T_m$.

The methods of pulsed heating of conductors and electrostatic levitation allow one to obtain properties in that high-temperature region,  unattainable for stationary studies. 

Pulse-heating involves high heating rates up to 10$^8$~K/s that leads to short measurement times (microseconds). This minimizes chemical interactions between the sample and its environment during the experiment. However, due to the high temperatures attained and the short duration of experiments, temperature measurements are carried out by optical methods. This requires the use of fast pyrometers and the knowledge of normal spectral emissivity as a function of temperature. Over the last decades temperature measurements on pulse heated liquid metal samples \cite{Bonnell:RU:1972, hixson1993thermophysical, boivineau1996thermophysical, obendrauf1993measurements, seifter2001microsecond} have always been performed under the assumption that emissivity is constant over the investigated temperature range of the liquid phase and its value equal to that at the melting point, due to the lack of data on the emissivity of liquid metals. This, in turn, can cause large deviations, especially at elevated temperatures, and leads to significant uncertainty in temperature dependent thermophysical properties.
Another potential weak point of this method is density measurements. Radial expansion measurements are made using the shadowgraph technique, so they can be affected by the surrounding medium. Since electrical resistivity measurements require the thermal expansion accounting, strong differences are often observed for experiments in liquid~\cite{Gathers:RPP:1986, Boivineau:IJMPT:2006}.

Electrostatic (ESL)~\cite{rhim1993electrostatic, ishikawa2005non, bradshaw2005machine, mauro2011highly, lee2013crystal, paradis2014materials}, aerodynamic (ADL)~\cite{millot2002high, wille2002thermophysical}, and electromagnetic (EML)~\cite{egry2000thermophysical} levitation methods are used to measure the density of liquid metals at high--temperature environment. Among them, only the ESL approach provides a quiescent condition and an almost spherical shape of the melt under high vacuum for measuring the density of a levitated liquid droplet. This allows easy image analysis for density, viscosity and surface tension measurements while still providing accurate results. However, discrepancies have been revealed between several reported density values, in particular for Fe~\cite{lee2013crystal}, Ti~\cite{lee2013crystal_Ti}, Zr~\cite{ishikawa2001new, ishikawa2003thermophysical}, and Nb~\cite{ishikawa2001new, ishikawa2003thermophysical}, in ESL experiments. Inaccurate liquid density values may lead to a misinterpretation in understanding of phase transitions and liquid structure as well as in predicting other physical quantities via numerical simulation.

Thus, the interpretation of both pulse-heating and levitation high-temperature experiments is difficult due to the speed and complexity of the occurring physical phenomena, therefore the experimental data obtained by different authors often differ significantly.

This is exactly the case with thermophysical properties of liquid Zr, which are still only approximately known.
%Thus, properties of zirconium in the liquid phase are still only known approximately. 
The density of liquid zirconium at the melting point measured by different authors differs within almost 10\%. The difference in the estimates of the slope of the thermal expansion curve is almost two-fold. Only few experimental measurements are available for temperatures above 2500~K.

In turn, theoretical description of zirconium in a condensed state is very challenging due to a complex electronic structure. In this connection, \textit{ab initio} methods based on density functional theory (DFT) which take into account the complicated electronic structure of transition metals are a great deal of help~\cite{Wimmer:JPCM:2010}. 

Properties of crystalline Zr are widely studied using DFT~\cite{Tu:NME:2018, Xin:JNM:2009, Xiong:JNM:2014, Khanal:NME:2021, Christensen:JNM:2015, Peng:JNM:2012, Willaime:JNM:2003, Hao:PRB:2008}. Works dedicated to molecular dynamics simulations are usually limited by using empirical interatomic potentials~\cite{Zhou:JNM:2018, Li:NME:2019}, although some of them are fitted to \textit{ab initio} data~\cite{Moore:JNM:2015}. On the other hand, computer simulations such as quantum molecular dynamics (QMD), that is based on DFT and does not use any empirical data except for fundamental physical constants, provide a physical insight into understanding of various phenomena on the atomic scale and enable one to predict some thermodynamic, mechanical, structural, transport and dynamic properties of materials, including in liquid or near-critical state. Such calculations are very computationally expensive and time consuming, so it is not surprising that only few QMD studies on Zr available~\cite{ Fang:CMS:2008, jakse2007short, Jakse:JNS:2007}.

Meanwhile, QMD was successfully implemented for the description of thermodynamical and transport properties of hot expanded molybdenum~\cite{Minakov:AIPADV:2018}, tungsten~\cite{Minakov:PRB:2018}, rhenium~\cite{Minakov:HTHP:2020}, boron~\cite{Clerouin:PRE:2008}, tantalum~\cite{Miljacic:CALPH:2015}, aluminum~\cite{Clerouin:PRB:2008}. It was shown that it is possible to describe consistently experiments both in the solid and in the liquid phase in the framework of \textit{ab initio} calculations. So QMD can be used as a reference method for the inspection of available experimental data.

The paper focuses on the thermophysical properties of both solid and liquid zirconium at high temperatures. The properties include density, pressure, temperature, molar enthalpy, resistivity, and normal spectral emissivity.
The authors believe that the new data obtained will help to improve the existing constitutive relations for zirconium, eliminating the uncertainty in its thermophysical properties at high temperatures and in the liquid state.

\section{Methods and simulation parameters}\label{sec:parameters}
\subsection{Thermodynamic properties}
The method of quantum molecular dynamics is used to obtain thermodynamic properties of zirconium. QMD simulations based on finite-temperature density functional theory (FT-DFT)~\cite{Mermin:PR:1965} are performed using the Vienna \textit{ab initio} Simulation Package (VASP)~\cite{Kresse:PRB:1993, Kresse:PRB:1994, Kresse:PRB:1996, Kresse:CMS:1996} with an implementation of the projector augmented-wave (PAW)~\cite{Blochl:PR:1994} pseudopotential. In all simulations the generalized gradient approximation (GGA) for the exchange-correlation (XC) functional with the Perdew-Burke-Ernzerhof (PBE)~\cite{Perdew:PRL:1996, Perdew:PRL:1997} parametrization is applied. To reduce numerical efforts, 12 valence electrons are included in our calculations. The Fermi smearing for electron occupancies is applied. To achieve convergence for both energy and pressure the plane wave cutoff energy is set to 400~eV. 250 atoms, initially arranged in the hcp structure for $\alpha$-phase and in the ideal bcc lattice for $\beta$-phase, with periodic boundary conditions are simulated. The liquid phase simulations start from a disordered state. The lattice parameter determines the density of matter $\rho$, which remains constant during the simulation process.
Nos{\'e}-Hoover~\cite{Nose:JCP:1984} thermostat is used to maintain a given temperature. The ion dynamics uses forces derived from the Hellmann-Feynman theorem, on the basis of the Born-Oppenheimer approximation. 
%; the ion dynamics is performed in the Born-Oppenheimer approximation with the canonical (NVT) ensemble. Then, by the Hellmann-Feynman theorem, the forces acting on the ions were computed and from the Newton’s equations new coordinates and velocities of the ions were obtained. 
%To reduce numerical efforts, the PAW potential with 12 valence electrons for 
The dynamics of Zr atoms is simulated within no less than 6~ps with 2~fs time step. In this paper we do not determine the melting temperature and use the reference value of $T_m = 2128$~K~\cite{CRC:2005}.

According to our previous work~\cite{Minakov:AIPADV:2018}, a study of the numerical convergence of QMD calculations on the number of atoms, \textbf{k}-points in the Brillouin zone and the estimation of statistical uncertainty in the averaging of thermodynamic properties are important aspects of accurate QMD modeling. Thus, we found that for an adequate description of the thermophysical properties of zirconium in the crystalline and liquid states, it is necessary to use 250 atoms in the supercell with the Baldereschi mean-value~\cite{Baldereschi:PRB:1973} \textbf{k}-point $\{ 1/4, 1/4, 1/4 \}$. Thermodynamic parameters of the system were found by averaging of the corresponding values at the equilibrium stage of the simulation. Since pressure strongly fluctuates during the calculation each thermodynamic state was simulated during no less than 3000 steps and the standard error of the calculated average pressure was estimated as 1~kbar. 
%The standard error of average energy estimated in a similar fashion is ??~kJ/mol.

The GGA-PBE XC functional gives the density of crystalline zirconium at normal conditions $\rho_0^{PBE} = 6.455$~g/cm$^3$ while the experimental value~\cite{CRC:2005} is $\rho_0^{exp} = 6.506$~g/cm$^3$; the difference is less than 1\%. 
%To obtain the equilibrium density under normal conditions for other pseudopotentials, the PHONOPY code\cite{togo2015first} for phonon calculations at quasi-harmonic levels was used. Local density approximation (LDA) with the Ceperley-Alder (CA) parametrization\cite{Ceperley:PRL:1980} gave an result: $\rho_0^{LDA} = ??$~g/cm$^3$, ??\% more than the experimental value. 
For comparison, cold density (at $T=0$ and $P=0$) for GGA-PBE is $\rho_{T=0}^{PBE} = 6.47$~g/cm$^3$, for AM05~\cite{Armiento:PRB:2005}  is $\rho_{T=0}^{AM} = 6.716$~g/cm$^3$, and for LDA-CA~\cite{Ceperley:PRL:1980} is $\rho_{T=0}^{CA} = 6.9$~g/cm$^3$.

\subsection{Transport properties}

The parallel GreeKuP code~\cite{Knyazev_2018} written by the authors is applied to determine the conductivity of Zr.
This code uses the matrix elements $\langle\Psi_i|\nabla_\alpha|\Psi_j\rangle$ from DFT calculation, to calculate the real part $\sigma_1(\omega)$ of the complex dynamic electrical conductivity $\sigma(\omega)=\sigma_1(\omega)+\rmi\sigma_2(\omega)$ according to the Kubo--Greenwood (KG)  formula~\cite{Desjarlais:PRE:2002,Knyazev2013,Knyazev_2018,Kowalski2007}:
%\begin{eqnarray*}
%  \dot x &=& v, 
%  \dot x &=& v, 
%  \dot v &=& F(x)
%\end{eqnarray*}
%\begin{eqnarray}
%2\times3&=&6\nonumber\\
%2+3&=&5\label{silly}
%\end{eqnarray}

%\begin{equation}
\begin{eqnarray}
\label{KuboGreenWood}
%\begin{gathered}
\sigma_1(\omega)=\frac{2\pi e^2\hbar^2}{3m_\mathrm e^2\omega\Omega}\sum_{i,j,\alpha,\mathbf k}\Big(W(\mathbf k)\big|\langle\Psi_{i,\mathbf k}|\nabla_\alpha|\Psi_{j,\mathbf k}\rangle\big|^2\times\nonumber\\
\times\big(f(\varepsilon_{i,\mathbf k})-f(\varepsilon_{j,\mathbf k})\big)
%\\
%\times
\delta(\varepsilon_{j,\mathbf k}-\varepsilon_{i,\mathbf k}-\hbar\omega)\Big).
\end{eqnarray}
%\end{gathered}
%\end{equation}
%$$i\rmi\mathrm i\mbox{i}$$
%is used for calculating of the real part of the dynamic electrical conductivity $\sigma_1(\omega)$ when $\omega\neq0$.
In this formula %$\sigma_1(\omega)$ can be treated as the energy absorption at the frequency $\omega$.
%Other designations:
$\Psi_{i,\mathbf k}$ and $\varepsilon_{i,\mathbf k}$ are the electronic eigenfunctions and eigenvalues, accordingly, for the $\textit{i}$th electronic band at a given $\mathbf k$-point in the Brillouin zone;
$W(\mathbf k)$ is the $\mathbf k$-point weight in the Brillouin zone using the Monkhorst-Pack scheme, $f(\varepsilon_{i,\mathbf k})$ is the Fermi distribution function;
$\Omega$ is the supercell volume, $e$ and $m_\mathrm e$ are the charge and mass of the electron, accordingly.
Formula~(\ref{KuboGreenWood}) takes into account all transitions between all possible states $i$ and $j$ to find $\sigma_1(\omega)$ which can be treated here as energy absorption at the corresponding frequency.
The $\delta$-function in the Kubo--Greenwood formula is responsible for the fact that only levels with the energy difference~$\hbar\omega$ make a contribution to the absorption at the frequency~$\omega$.
If the occupation of an initial level is equal to the occupation of a last level, then the probability of the transition between these levels is 0, the energy absorption is 0 too, so the difference of occupation numbers $\big(f(\varepsilon_{i,\mathbf k})-f(\varepsilon_{j,\mathbf k})\big)$ is included to this formula. The $\delta$-function is represented by a Gaussian with broadening of $\Delta = 0.05$~eV in our calculations. 
The squared matrix elements $|\langle\Psi_{i,\mathbf k}|\nabla_\alpha|\Psi_{j,\mathbf k}\rangle|^2$ are responsible for the intensity of every such transition ($\nabla_\alpha$ is the velocity operator along the spatial direction $\alpha$, there are three such spatial directions in total).

The value of dc conductivity can be obtained by extrapolation of the dynamic conductivity to zero frequency~\cite{Knyazev2013}. The convergence upon width of the $\delta$-function is carefully checked for every calculation.

The conductivity calculation requires more precise DFT determination of electronic structure and consequently it is significantly more computationally expensive. Therefore, such calculations are performed not for all but selected configurations from the equilibrium QMD trajectory. The averaging of the calculated conductivity is performed over no less than 10 configurations equidistantly spaced along the ions trajectory in our QMD+KG calculations.
We use $2 \times 2 \times 2$ Monkhorst--Pack \textbf{k}-points grid~\cite{Monkhorst:PRB:1976} in the Brillouin zone for this type of calculations. 

%At this stage we use other values of some technical parameters (finer \textbf{k}-points grid, more number of bands) which can provide better accuracy compared with the calculation of thermodynamic properties.

\subsection{Optic properties}

The real and imaginary parts of a complex function $\sigma(\omega)=\sigma_1(\omega)+\rmi\sigma_2(\omega)$ that is analytic in the upper half-plane, are connected via the Kramers--Kronig transform (KKT): if we know the real part $\sigma_1(\omega)$ we can use the relation
%$$\sigma_1(\omega)=\frac{1}{\pi}\mathcal{P}\int_{-\infty}^{\infty}\frac{\sigma_2(\omega^\prime)}{\omega^\prime-\omega}d\omega^\prime$$
%to find real part and the relation
\begin{equation}\label{KrKrTr-Im}
%\sigma_2(\omega)=-\frac{1}{\pi}\mathcal{P}\int_{-\infty}^{\infty}\frac{\sigma_1(\omega^\prime)}{\omega^\prime-\omega}d\omega^\prime
\sigma_2(\omega)=-\frac{2\omega}\pi\mathcal P\int_0^\infty\frac{\sigma_1(\omega^\prime)}{\omega^{\prime2}-\omega^2}d\omega^\prime
\end{equation}
%\textcolor[rgb]{1.0.0}{(poep opyy ec koe!)}
to find the imaginary part $\sigma_2(\omega)$ of this function, where $\mathcal{P}$ denotes the Cauchy principal value.
For any stable physical system, causality implies the condition of analyticity, and conversely, analyticity implies causality of the corresponding stable physical system~\cite{Toll_1956}.

At known $\sigma(\omega)=\sigma_1(\omega)+i\sigma_2(\omega)$ we can obtain optical properties depending on the radiation frequency $\omega$:
\begin{itemize}
	\item
	the complex dielectric constant $\varepsilon(\omega)=\varepsilon_1(\omega)+\rmi\varepsilon_2(\omega)$ as
	\begin{equation*}
	%2\cdot2
	\varepsilon_1(\omega)=1-\frac{\sigma_2(\omega)}{\omega\varepsilon_0},\qquad\varepsilon_2(\omega)=\frac{\sigma_1(\omega)}{\omega\varepsilon_0},
	\end{equation*}
	where $\varepsilon_0$ is the vacuum permittivity;
	\item
	the complex refractive index $n(\omega)+\rmi k(\omega)$ as
	\begin{equation*}
	n(\omega)=\sqrt{\frac{\big|\varepsilon(\omega)\big|+\varepsilon_1(\omega)}{2}},\qquad k(\omega)=\sqrt{\frac{\big|\varepsilon(\omega)\big|-\varepsilon_1(\omega)}{2}};
	\end{equation*}
	\item
	the normal spectral reflectivity as
	\begin{equation*}
	R(\omega)=\frac{\big(1-n(\omega)\big)^2+k(\omega)^2}{\big(1+n(\omega)\big)^2+k(\omega)^2};
	\end{equation*}
	\item
	the absorption coefficient as
	\begin{equation*}
	\alpha(\omega)=2k(\omega)\frac{\omega}{c},
	\end{equation*}
	where $c$ is the speed of light in vacuum;
	\item
	the normal spectral emissivity as
	\begin{equation}
	\label{eq:emissivity}
	\mathcal{E}(\omega)=1-R(\omega).
	\end{equation}
\end{itemize}

%The function $\sigma_1(\omega)$ is given as a one--dimension array with different values $\sigma_i$ for different values $\omega_i$ with a step $\Delta\omega$. As $\sigma_1(\omega)$ may be a non--differentiable function we use the trapezoidal rule for integration.
%In this program we can set a different upper integration limit $\omega_\mathrm{max}$ and change the integration step $\Delta\omega$.
%The cubic spline of the original function $\sigma_1(\omega)$ is used in case of a small integration step $\Delta\omega$ (implemented using \verb|GSL| library).
%A user can also increase the upper integration limit $\omega_\mathrm{max}$ beyond the maximum frequency value of the array.
%In this case, additional elements of $\sigma_1(\omega)$ array are populated with the last value of the original array.
%Convergence according to the parameter $\eta$ was tested and achieved at $\eta\le10^{-6}$.

The integration in practical KKT calculation is performed over the limited frequency range. In this work up to 8000 orbitals were involved in the DFT calculation with about 6000 unoccupied ones to account for high--energy transitions in the Kubo--Greenwood formula~(\ref{KuboGreenWood}). We estimate the computational error of our calculation of optic properties of less than 1\%. 

\section{Results and Discussion}
\subsection{Density}
\begin{figure}[t]
\includegraphics[width=0.99\columnwidth]{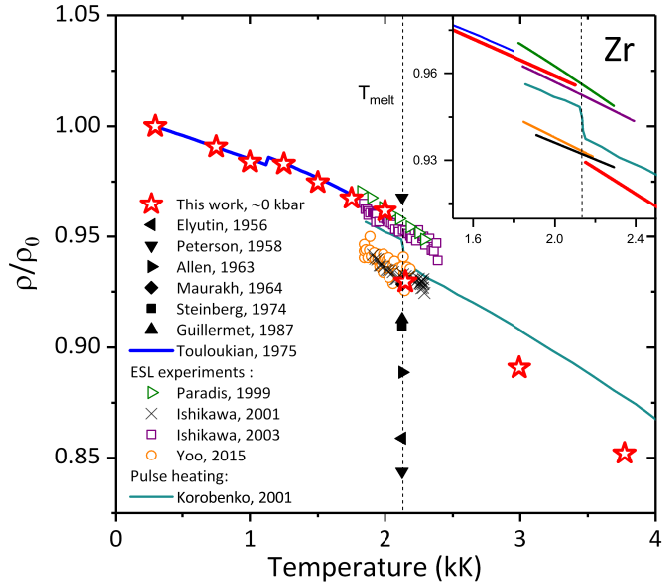}
\caption{Isobaric expansion of solid $\alpha$ and $\beta$-Zr as well as liquid Zr in the relative density versus temperature diagram. Star symbols are QMD data. Experimental data: measurements of Zr density at the melting point~\cite{peterson1958surface, allen1963surface, yelyutin1964viscosity, steinberg1974simple, guillermet1987critical} are shown by filled symbols; electrostatic levitation data~\cite{paradis1999thermophysical, ishikawa2003thermophysical, yoo2015uncertainty} are shown by empty symbols and data by Ishikawa~\cite{Ishikawa:RSI:2001} are shown by diagonal crosses; pulse-heating measurements~\cite{korobenko2001temperature} is a turquoise line. The inset is a closer look at the melting region; data are represented by their approximation curves, colors are the same as corresponding symbol colors. 
}
\label{fig:rho-t}
\end{figure}

\begin{figure}[t]
\includegraphics[width=0.99\columnwidth]{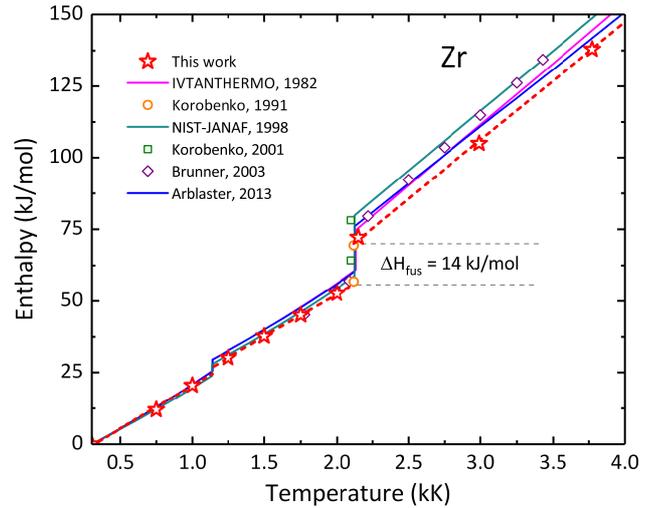}
\caption{Molar enthalpy versus temperature for Zr. Star symbols are QMD data. Linear fits for QMD points in the ranges 300--1000, 1150--1750 and 2150--3770~K are shown by red dashed lines and extrapolated to higher and lower temperatures. The turquoise line is a relation from Thermochemical Table NIST~\cite{chase1998nist}, blue line is an approximation from Arblaster \cite{arblaster2013thermodynamic} overview, pink line is an approximation by IVTANTHERMO Handbook~\cite{Gurvich:IVTANTHERMO:1982}. Other experimental data~\cite{korobenko2001experimental, brunner2003normal, korobenko1991properties} are shown by symbols.
% and data of Korobenko 1991~\cite{korobenko1991properties} are shown by crosses.  
}
\label{fig:enthalpy-temp}
\end{figure}

\begin{figure}[t]
\includegraphics[width=0.99\columnwidth]{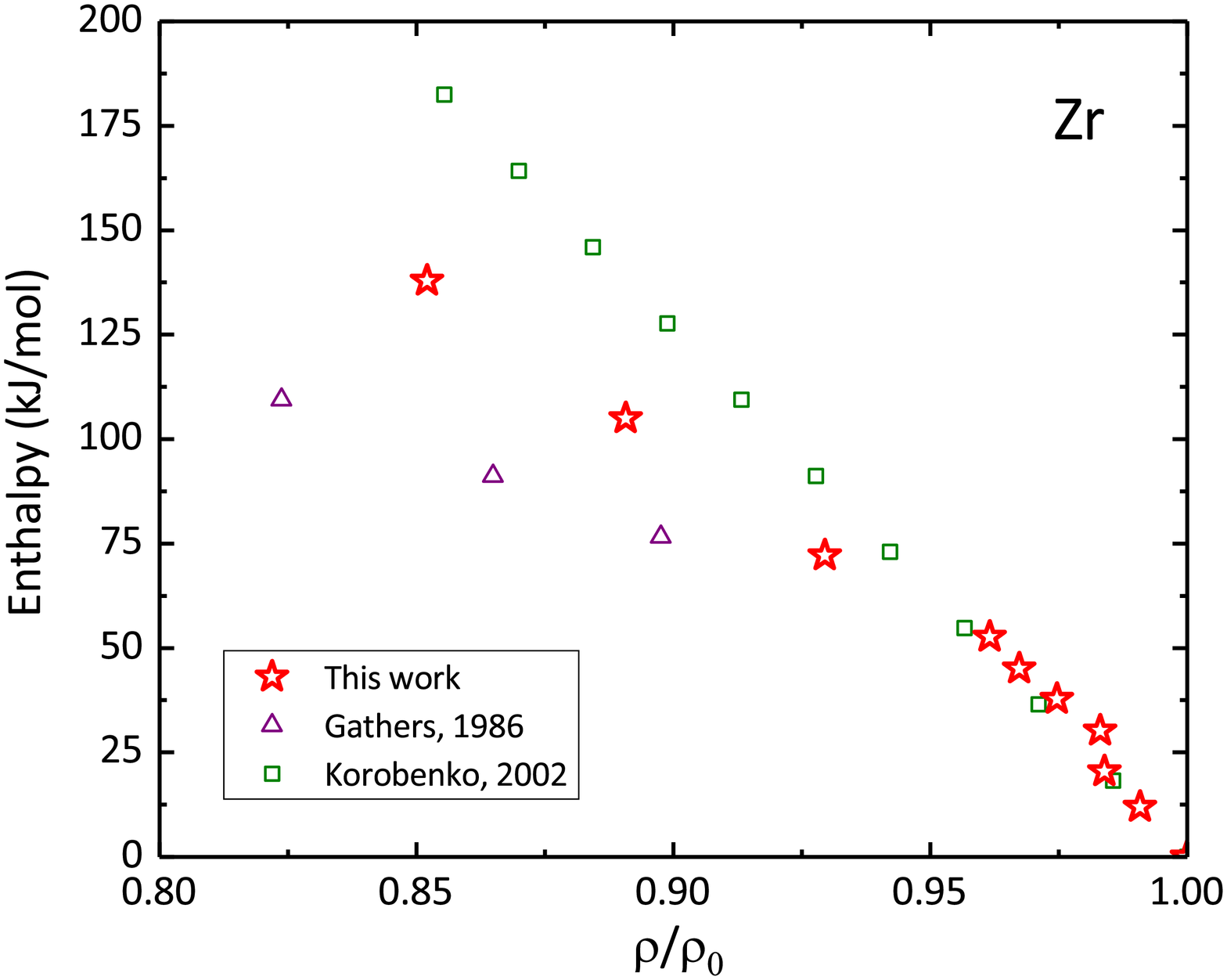}
\caption{Molar enthalpy versus relative density for Zr. Star symbols are QMD data. Pulse-heating measurements~\cite{korobenko2002zirconium, gathers1986dynamic} are shown by empty symbols. 
}
\label{fig:enthalpy-dens}
\end{figure}

The main source of experimental data on thermophysical properties of metals in the liquid phase at high temperatures near the binodal is the dynamic pulse-heating method. The obtained quantities are enthalpy, electrical resistance, volume expansion, specific heat capacity, thermal conductivity, and some others.

%It is a rapid heating of a metal (a thin wire specimens several centimeters long) by an electrical current pulse. During the period of heating, the wire expands in the liquid phase laterally. Heating current and specimen voltage drop are recorded using either oscilloscopes. Temperature measurements are performed using an optical pyrometer. From the obtained values, the enthalpy, electrical resistance, volume expansion, specific heat capacity, thermal conductivity, and some others are calculated.

Electrostatic levitation (ESL) technique~\cite{paradis1999thermophysical} allows a sample to levitate in a high vacuum and to process it at high temperature using laser radiation. Thus, a sample is isolated from contaminating container walls as well as surrounding gases. However, measurements are limited by the temperatures of about 2500~K. 
%This method uses the Coulomb force between a charged sample and the electrodes, and the position of the sample is controlled using high-speed feedback from the camera image. 
%The sample was heated by the radiation coming from two CO2 lasers. 

An experimental pressure value during isobaric expansion (IEX) measurements is usually comparable to the absolute computational error of the calculation due to pressure fluctuations during the simulation. Increasing the number of atoms in the simulation to reduce oscillations leads to a non-linear increase in computational costs~\cite{Hafner:JCC:2008, Harl:PRL:2009}. Therefore, in order to reduce the influence of calculation errors and to be able to reconstruct the IEX curve at a given pressure with high accuracy, we perform a series of calculations along isochores in the liquid phase and along isotherms in the solid phase. Then calculated points are interpolated by a quadratic polynomial on isochores and a linear fit on isotherms, and interpolation relations are used to restore the isobar.
%The same scheme is used for reconstruction of dc conductivity along the isobar. 
%The calculated points on isotherms for the solid state and on isochores for the liquid state are collected in the Table~\ref{tab:eos} of  Appendix~\ref{app:eos}.

Figure~\ref{fig:rho-t} shows our calculations and available experimental data on isobaric expansion of zirconium in the melting region. The results of density calculations and measurements are compared in relative units, since the calculated density under normal conditions differs somewhat from the experimental value. We use the QMD-calculated reference density at normal conditions $\rho_0 \equiv \rho_0^{PBE} = 6.455$~g/cm$^3$ for our data.

As can be seen, calculations in solid $\alpha$ and $\beta$ phases agree perfectly with the reference data of the Handbook by Touloukian \textit{et al.}~\cite{touloukian1975thermophysical}, which provides a compilation of multiple experiments for solid Zr up to 1800~K.
% which are semiempirical relationship between coefficient of thermal expansion and melting temperature.
At higher temperatures data of ESL measurements for $\beta$-Zr are available.
% Also the slope of the thermal expansion curves between them and 
 The slope of our curve corresponds to the data by Ishikawa \textit{et al.}~\cite{ishikawa2003thermophysical}, obtained using the electrostatic levitation furnace and validated with Hard Sphere model~\cite{itami1984application} calculations. 
 %As Ishikawa points out, drop oscillation might be damped by the electrical field used to levitate samples against gravity, leading to small uncertainties. %Ishikawa2003(measured by containerless methods and compared with calculations based on the HS model). Moreover, drop oscillation might be damped by the electrical field used to levitate samples against gravity. For this reason, experiments performed under microgravity are necessary to reduce the uncertainty.
Our calculations confirm the density of Zr at the melting temperature as measured by electrostatic levitation by Ishikawa \textit{et al.}~\cite{ishikawa2003thermophysical}, as well as earlier measurements by Paradis and Rhim~\cite{paradis1999thermophysical}.
% but are inconsistent with the higher density and temperature of Tolukyan.

Experimental data on density of molten Zr from the following earlier papers are shown in Fig.~\ref{fig:rho-t}: Elyutin, Peterson \textit{et al.}~\cite{peterson1958surface}, Allen~\cite{allen1963surface}, Maurakh \textit{et al.}~\cite{yelyutin1964viscosity}, Steinberg~\cite{steinberg1974simple} and Guillermet and Fern{\'a}ndez~\cite{guillermet1987critical}. As can be seen from the figure, the data spread is very significant. Noteworthy is the huge density drop from 6.31 to 5.5~g/cm$^3$ at melting, reported by Peterson \textit{et al.}

Meanwhile, the density of liquid Zr at the melting point by QMD is in good agreement with the results of electrostatic levitation experiments by Ishikawa \textit{et al.}~\cite{Ishikawa:RSI:2001} and Yoo \textit{et al.}~\cite{yoo2015uncertainty} where Zr was in supercooled and stable liquid phases.
As can be seen, the most recent experimental data by Yoo \textit{et al.} have some scatter and the error in measuring the density of liquid zirconium was estimated as $\pm$1.1$\%$. Nevertheless, the average slope of experimental data~\cite{yoo2015uncertainty} is consistent with our slope of liquid Zr thermal expansion.

Pulse-heating experiment by Korobenko and Savvatimskii~\cite{korobenko2001temperature} is the only high-temperature density--temperature measurement available for Zr. The QMD slope of the thermal expansion curve in the solid phase is consistent with this experiment, but there is a deviation in the liquid phase. The density of solid Zr at the melting point in this experiment is lower than in the previously discussed works, but the density of liquid is the highest. QMD also predicts lower density of liquid at the same temperature.
Bending deformations which are not detected in thin shadowgraph images of the wire diameter could be reasonable explanation of the underestimation of liquid volume in the experiment. However, the authors points out that bending deformations are insignificant under experimental conditions and for the given rate of heating.

\subsection{Enthalpy}

Another thermophysical parameter available in dynamic heating experiments is enthalpy. As temperature and enthalpy can be measured simultaneously as a function of time, enthalpy as a function of temperature can be easily obtained~\cite{Gathers:RPP:1986, Boivineau:IJMPT:2006}. The enthalpy change along the zero isobar $\Delta H \equiv H_T - H_{300} = U_T - U_{300}$, is considered between a given temperature and $T=300$~K directly from QMD calculations where $U_T$ and $U_{300}$ are the corresponding internal energies of the system. To reduce the systematic error the convergence of $U_{300}$ was carefully tested (see Table~\ref{tab:conv} in \ref{app:convergence}).

The results of our QMD calculations on molar enthalpy depending on temperature are presented in Fig.~\ref{fig:enthalpy-temp} together with reference and experimental data. 

%As can be seen from the figure, the discrepancy between the experimental data in the solid phase is noticeable only in the region of the $\alpha$--$\beta$ transition. But the points deviating from the rest of the data belong to the works of seventy years ago by Redmond and Lones~\cite{redmond1952enthalpies} and Coughlin and King~\cite{coughlin1950high}. 
%In the first work, measurements of the relative enthalpy were made using an ice calorimeter\cite{redmond1951design}. To do this, the capsuled sample was heated in a furnace to a constant temperature, and was then dropped into the calorimeter where the total enthalpy of the capsule and its contents, referred to 273~K (temperature of calorimeter), was measured. To obtain the enthalpy of the sample alone the enthalpy of the empty capsule was subtracted. The enthalpy of the empty capsule was determined in the same manner as the enthalpy of the filled capsule. In the second work, measurements were carried out by the "dropping" method using an apparatus\cite{southard1941modified} for heat contents at high temperatures (to 1500~$^{\circ}$K), in which a capsule containing the sample is heated in a furnace to a determined temperature and at a given moment dropped into a calorimeter of known heat capacity. 

In a relatively recent review on zirconium, Arblaster~\cite{arblaster2013thermodynamic} provides empirical relationships between enthalpy and temperature based on experimental data for the $\alpha$, $\beta$, and liquid phases. Above 273.15~K and up to the $\alpha$--$\beta$ transition temperature, the Douglas and Victor~\cite{douglas1958heat} enthalpy measurements (373--1124~K) were adopted after a correction to the International Temperature Scale of 1990 (ITS-90 is an approximation of the thermodynamic temperature scale that facilitates comparability and compatibility of temperature measurements internationally). Before melting, specific heat measurements of Cezairliyan and Righini~\cite{cezairliyan1974simultaneous} (1500--2100K) and Petrova \textit{et al.}~\cite{petrova2000investigation} (1200--2100~K) were combined with the drop calorimetry enthalpy measurements of Rösner-Kuhnetal \textit{et al.}~\cite{rosner2001enthalpy} (1821--2105~K).

The IVTANTHERMO Handbook~\cite{Gurvich:IVTANTHERMO:1982, Belov:JPCS:2018} data for solid Zr is almost coincide with those of Arblaster.

%In the $\beta$-phase, the electrical pulse heating performed by Brunner~\cite{brunner2003normal}, combining a fast laser polarimetry technique with setup for high-speed measurements on liquid metal samples at high temperatures corresponds to the Arblaster dependence. A laser polarimeter was used for the determination of optical parameters, because the normal spectral emissivity is an important quantity for temperature determination when measuring thermophysical properties.

Thermochemical Table NIST~\cite{chase1998nist} provides an approximation of seven enthalpy studies~\cite{jaeger1934exact, coughlin1950high, skinner1951vapor, adenstedt1952physical, redmond1952enthalpies, douglas1958heat, fieldhouse1961armour} between 300~K and 1135~K and four~\cite{coughlin1950high, skinner1951vapor, redmond1952enthalpies, douglas1958heat} over the temperature range 1135--2125~K.

Our calculations agree excellently with all reference data in the $\alpha$-phase, while for $\beta$-Zr better agreement is observed between our data and NIST approximation and measurements of Brunner \textit{et al.}~\cite{brunner2003normal}.

At the melting point, there are two datasets obtained by the electrical pulse current heating method by Korobenko and his colleagues. In later work~\cite{korobenko2001experimental} with a heating rate of $3 \times 10^8$~K~s$^{-1}$, the molar enthalpy of solid and liquid zirconium was reported as 64.1 and 78.1~kJ/mol, respectively, while in earlier work~\cite{korobenko1991properties} with a heating rate of $2 \times 10^7$~K~s$^{-1}$, the molar enthalpy of solid and liquid zirconium were 56.8 and 69.4~kJ/mol, correspondingly, that is in excellent agreement with our results.
It is worth to note, that in Ref.~\cite{korobenko2001experimental} experiments with zirconium foils were carried out, while zirconium wires were used in Ref.~\cite{korobenko1991properties} .

In the liquid phase of zirconium, the difference between the experimental and reference enthalpy data and the QMD results is more noticeable. In Thermochemical Table NIST~\cite{chase1998nist} there are no enthalpy studies covering the liquid region; a constant value of $C_{p} = 41.84$~J K$^{-1}$mol$^{-1}$ is assumed for the region 2125--5500~K. In the review by Arblaster~\cite{arblaster2013thermodynamic} the drop calorimetry enthalpy measurements of Bonnell \cite{Bonnell:RU:1972} (2232--3046~K), Kats \textit{et al.} \cite{kats1985proprietes} and Qin \textit{et al.} \cite{qin1997spectral} (2027--2901~K) were separately fitted to linear functions and combined for the liquid phase. Brunner's \textit{et al.} data~\cite{brunner2003normal} on electrical pulsed heating intersects all reference extrapolated lines.

As can be seen from the figure, the slope of our curve in the liquid phase is similar to the reference ones. The isobaric heat capacity in the liquid Zr by QMD is  40.5~J~K$^{-1}$mol$^{-1}$, while Arblaster presented the value of 39.92, IVTANTHERMO~--- 42.6, and NIST~--- 41.84~~J~K$^{-1}$mol$^{-1}$. 

Thus, the difference in the molar enthalpy in the liquid phase is caused by the heat of fusion estimate.
We obtained a value of heat of fusion of zirconium $\Delta H_{fus} \approx$~14~kJ/mol by QMD. 
Our calculations tend to the lower values of the enthalpy of melting measured by rapid pulse heating~\cite{korobenko2001experimental, korobenko1991properties, brunner2003normal} and drop calorimetry~\cite{elyutin1967determining, kats1985proprietes, rosner2001enthalpy} (12.8--21.5~kJ/mol).

The diagram of molar enthalpy versus relative density is shown in Fig.~\ref{fig:enthalpy-dens}. Only data on electrical pulse heating obtained by Gathers~\cite{gathers1986dynamic} and Korobenko \textit{et al.}~\cite{korobenko2002zirconium} are shown. The available experimental data notably contradict to each other and our QMD calculations in the liquid phase. But in the solid there is good agreement between our data and Korobenko's measurements.

%Slope QMD: $-0.31 \times 10^{-3}$~g/(cm$^3$K)
%
%			Paradis: $-0.30 \times 10^{-3}$~g/(cm$^3$K)
%
%			Ishikawa 2001: $-0.19 \times 10^{-3}$~g/(cm$^3$K)
%
%			Ishikawa 2003: $-0.22 \times 10^{-3}$~g/(cm$^3$K)
%			
%			Yoo: $-0.23 \times 10^{-3}$~g/(cm$^3$K)
%
%			Korobenko: $-0.24 \times 10^{-3}$~g/(cm$^3$K)

\subsection{Resistivity}
			
\begin{figure}[t]
\includegraphics[width=0.99\columnwidth]{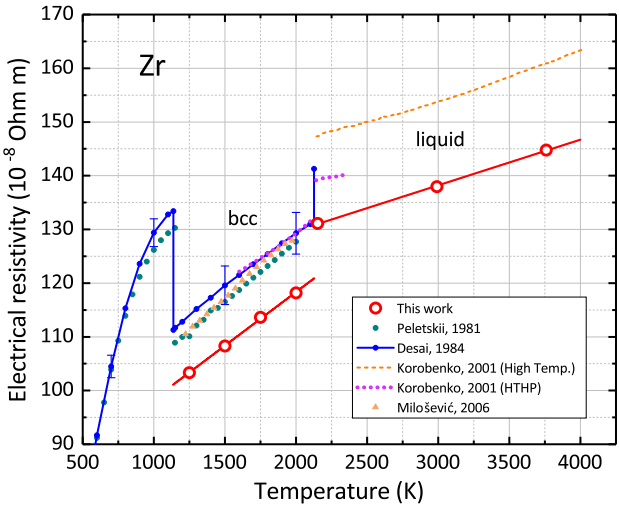}
\caption{Electrical resistivity of Zr near melting. QMD data are red open circles, red lines are linear approximation. Recommended values by Desai~\cite{Desai:JPCRD:1984} are shown as blue line with dots. Recommended value by Peletskii~\textit{et al.}~\cite{Peletskii:ME:1981} are shown as turquoise dots. Pulse-heating measurements by Korobenko~\textit{et al.}~\cite{korobenko2001temperature, korobenko2001experimental} are shown by dashed orange and dotted purple lines, correspondingly. Subsecond pulse calorimetry by Milo\v{s}evi\'{c}~\cite{Milosevic:IJT:2006} is orange triangles.}
\label{fig:resist-t}
\end{figure}		
\begin{figure}[t]
\includegraphics[width=0.99\columnwidth]{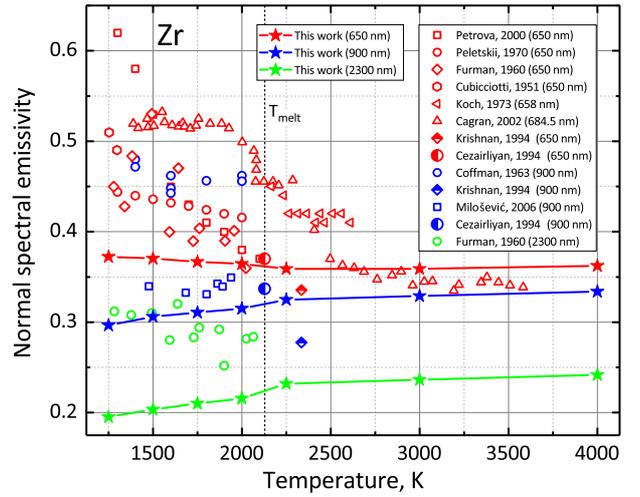}
\caption{Normal spectral emissivity versus temperature. Star symbols are QMD data. Other symbols are experimental data~\cite{Petrova:HT:2000, Peletskii:TVT:1970, Furman:1960, Cubicciotti:JACS:1951, Koch::1973, Cagran:HTP:2002, Krishnan_1994, Cezairliyan:IJT:1994, Coffman:P:1963, Milosevic:IJT:2006}, colored by wavelength under study. Red symbols are 650~nm and close values, blue symbols are 900~nm, green symbols are 2300~nm.
}
\label{fig:emis-t}
\end{figure}	

Electrical resistivity of solid zirconium was widely investigated in numerous experiments.
Experimental data on resistivity of solid zirconium in the range of temperature 1--2127~K were collected by Desai~\cite{Desai:JPCRD:1984} in 1984. An extensive overview provides a compilation of 43 experimental data sets. Earlier in 1981, recommended data for Zr resistivity were published in the Handbook by Peletskii and Belskaya~\cite{Peletskii:ME:1981}. Meanwhile, data on resistivity of liquid Zr are almost lacking. Experiments on the electrical explosion of zirconium foils and wires by Korobenko \textit{et al.}~\cite{korobenko2001experimental,korobenko2001temperature,korobenko2002zirconium} made it possible to obtain resistivity of liquid zirconium up to 4100~K.
The data discussed are shown in Fig.~\ref{fig:resist-t}, as well as recent experimental data for solid Zr by Milo\v{s}evi\'{c}~\textit{et al.}~\cite{Milosevic:IJT:2006}. As can be seen, there is a consensus on the resistivity of $\beta$-Zr. The jump in resistivity due to melting differs more than two-fold even by the same authors.

We found that QMD+KG provides systematically lower resistivity for Zr than observed in experiments. Meanwhile, the slope of the resistivity for $\beta$-Zr is in excellent agreement with Desai data. The difference between our data and Desai approximation is about 11~$\mu\Omega$cm, or about 10\%. The jump in resistivity at melting by QMD agree perfectly with Desai estimate as well. The slope of the liquid part of the QMD curve of resistivity is consistent with one by Korobenko~\textit{et al.}~\cite{korobenko2001temperature}.

Discussing the discrepancy between our results and experiments, we firstly note, that the statistical error is less than 1\% in our calculations and the convergence on simulation parameters was well checked, see \ref{app:convergence}.
 Meanwhile, it should be noted, that QMD+KG calculation of dc conductivity is sensitive to the choice of pseudopotential~\cite{French:PRB:2014}. We used the most accurate pseudopotential that is available in the VASP package for Zr. Among the disadvantages of the QMD+KG method we can also mention an underestimation of contributions from electron–electron collisions~\cite{Petrov:JL:2016, Desjarlais:PRE:2017}, which, however, seem to have little effect at the conditions considered in this paper. Finally, it should be remembered, that the DFT electronic band structure is based on Kohn--Sham eigenvalues that can differ from a real system. For example, DFT is assumed to work good for metals, but underestimates the size of the band gap in insulators and semiconductors to varying degree depending on the material. This phenomenon may have a systematic effect on the calculated dc conductivities~\cite{Desjarlais:PRE:2002}. Thus, comparison with well--established experimental data is still required for the resistivity calculations using QMG+KG in order to estimate the accuracy of calculated quantities or to envisage a corresponding correction.

At the same time, it should be mentioned here, that the purity of Zr samples has notable influence on the results of resistivity measurements. According to Peletskii and Belskaya~\cite{Peletskii:ME:1981} less impurities in Zr samples provide a decrease in measured resistivity.

%Desai \cite{Desai:JPCRD:1984}
%Increasing purity leads to decrease in resistivity.
%Handbook by Peletskii and Belskaya \cite{Peletskii:ME:1981}
%Milosevic \cite{Milosevic:IJT:2006}

\subsection{Normal spectral emissivity}
\begin{figure}[t]
\includegraphics[width=0.99\columnwidth]{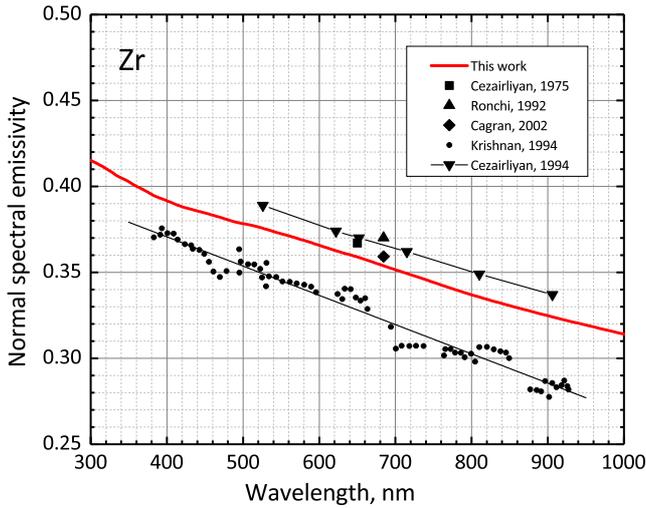}
\caption{Normal spectral emissivity of liquid Zr at different wavelengths. Data obtained in this work is shown as red line; black circles are experiment~\cite{Krishnan_1994}, black line is its linear approximation; black square is experiment~\cite{Cezairliyan1975}, black triangle is experiment~\cite{Ronchi_1992}, black diamond is experiment~\cite{Cagran:HTP:2002}, down triangles with line is experiment~\cite{Cezairliyan:IJT:1994}.}
\label{fig:emis-wave}
\end{figure}

Optical properties of solid zirconium were studied in multiple experiments~\cite{Peletskii:TVT:1970, Baria:MT:1974, Cezairliyan:JRNBSSPC:1974, Petrova:HT:2000, Milosevic:IJT:2006}. Most of the experimental data for the liquid phase are close to the melting temperature~\cite{Cezairliyan1975, Cagran:HTP:2002, brunner2003normal, Krishnan_1994, Ronchi_1992}.
Discussing the optical properties, we will focus on normal spectral emissivity.

Figure~\ref{fig:emis-t} demonstrates some available experimental data on normal spectral emissivity of Zr as well as our calculated data for several wavelengths in the range of temperatures 1200--4000~K. 

The scatter of the data for solid zirconium is particularly conspicuous. The detailed explanation of this fact was presented by Cagran~\textit{et al.}~\cite{Cagran:HTP:2002}.
The fact is, emissivity is mainly influenced by the roughness of the sample surface and possible surface impurities such as oxide layers. 
So a large variety of normal spectral emissivity values for the solid state can be explained by different procedures of preparing and abrading the sample.
%generating arbitrary results for the solid state, which are not influenced by the specimen material itself, but mostly by the produced surface structure.
To obtain reproducible values for the solid state, one must preheat the sample several times close to melting to smoothen the surface and to eliminate surface contaminations. However, this procedure is usually overlooked by researchers.

Again, some experiments exhibit a sudden drop in normal spectral emissivity at the onset of melting (noticeable for Cagran data in Fig.~\ref{fig:emis-t}).
This is believed to be related to a smoothening of the sample surface upon melting and to the evaporation of possible oxide layers from the surface. So during the melting transition, emissivity drops to a lower, reproducible, and material-specific value.
Moreover, some recent measurements on copper, silver, and platinum showed that the opposite behaviour is equally possible, when normal spectral emissivity suddenly increases at the start of melting. This occurs when the surface of the sample is almost perfectly smooth and does not need to be polished~\cite{Cagran:HTP:2002}.

The QMD+KG calculated normal spectral emissivity for Zr at $\lambda = 650$~nm does not demonstrate an abrupt drop at melting. Moreover, for larger wavelengths an opposite effect is predicted by our calculations. Our results for 650 and 900~nm agree almost perfectly with Cezairliyan~\textit{et al.}~\cite{Cezairliyan:IJT:1994} experiment for molten Zr. There is good agreement of our data with Cagran measurements for the liquid Zr and with recent experiment by Milo\v{s}evi\'{c}~\textit{et al.} for $\beta$-Zr. Finally, our calculations predict almost constant normal spectral emissivity up to 4000~K and do not confirm the growth of $\mathcal{E}$ after 3000~K, observed in some experiments~\cite{Savvatimskii:EI:2012}.

The comparison of the calculated normal spectral emissivity of liquid Zr with experimental data in the vicinity of melting temperature is also shown in Figure~\ref{fig:emis-wave}.
We can see that the slope of $\mathcal{E}(\lambda)$ obtained in this work is similar to the linear fit of the experimental data by Krishnan~\textit{et al.}\cite{Krishnan_1994}. Even more spectacular agreement is observed for data of Cezairliyan~\textit{et al.}~\cite{Cezairliyan:IJT:1994}. The values of normal spectral emissivity from first-principle calculations agree well with other experiments for $\lambda=650$~\cite{Cezairliyan1975} and 684.5~nm~\cite{Ronchi_1992,Cagran:HTP:2002}.

\section{Conclusion} 

Thermophysical properties of zirconium in the vicinity of melting are very ambiguous due to strong contradictions in the available experimental data, that makes it difficult to construct reliable equation of state of Zr. However, in this work we have demonstrated for the first time that some experimental data can be accurately described using the \textit{ab initio} method of quantum molecular dynamics. 

In the solid phase the results on thermal expansion by QMD calculations agree perfectly with the approximation from Tolukyan~\textit{et al.} Handbook~\cite{touloukian1975thermophysical} and electrostatic levitation measurements by Ishikawa \textit{et al.}~\cite{ishikawa2003thermophysical}.
%, the slope of the thermal expansion curves of which also coincides. 

Thermal expansion of liquid Zr by QMD is in good agreement with the results of experiments on electrostatic levitation by Ishikawa \textit{et al.}~\cite{Ishikawa:RSI:2001} and Yoo \textit{et al.}~\cite{yoo2015uncertainty}. Meanwhile, the QMD slope of the thermal expansion curve in the liquid phase is not consistent with the high-temperature pulse-heating experiment by Korobenko and Savvatimskii~\cite{korobenko2001temperature}.

Our QMD calculations predict a substantial volume change on melting of about $2.6\%$, in addition to quite a low value of calculated enthalpy of fusion (${\approx}\,14$~kJ/mol). This fact leads to the steep slope of the melting curve at $P\,{=}\,0$ of $dT_m/dP\,{=}\,60$~K/GPa predicted by QMD contrary to laser-heated diamond anvil cell experiments~\cite{parisiades2019melting, Radousky:PRR:2020, Pigott:JPCM:2020} where 24--28~K/GPa was proposed.

However, it should be noted, that the classic molecular dynamics with empirical interatomic EAM potential predicts an initially steep melting curve for Zr ($dT_m/dP\,{=}\,64$~K/GPa at $P\,{=}\,0$ GPa) that quickly flattens out at higher pressures~\cite{Smirnova:CMS:2018}. Moreover, recently similar steep behavior of the melting curve at low pressures was confirmed in experiments for Ti~\cite{Stutzmann:PRB:2015}. 
 
We found that QMD+KG systematically underestimates resistivity for Zr compared to experiments. Meanwhile, the slope of the resistivity for $\beta$-Zr and a jump in resistivity at melting is in excellent agreement with Desai approximation~\cite{Desai:JPCRD:1984}. The slope of temperature dependence of resistivity in liquid Zr is consistent with one by Korobenko~\textit{et al.}~\cite{korobenko2001temperature}.

Contrary to some experiments, the calculated normal spectral emissivity for Zr at $\lambda\,{=}\,650$~nm does not demonstrate an abrupt drop at melting. Moreover, for larger wavelengths the jump in normal spectral emissivity is predicted by our calculations. Our results for 650 and 900~nm agree excellently with Cezairliyan~\textit{et al.}~\cite{Cezairliyan:IJT:1994} and some other experiments for molten Zr. Good agreement of our data with Cagran~\textit{et al.}~\cite{Cagran:HTP:2002} for the liquid Zr and with recent experiment by Milo\v{s}evi\'{c}~\textit{et al.} for $\beta$-Zr is also demonstrated. Our calculations predict almost constant value of normal spectral emissivity up to 4000~K.

\section*{Acknowledgements}
This work has been supported by the Russian Science Foundation (grant No.\,20-79-10398). The authors acknowledge the JIHT RAS Supercomputer Centre, the Joint Supercomputer Centre of the Russian Academy of Sciences, and the Shared Resource Centre “Far Eastern Computing Resource” IACP FEB RAS for providing computing time.

\appendix
\section{\label{app:convergence}Convergence}

Table~\ref{tab:conv} contains convergence tests for calculated thermodynamic properties upon system size and number of \textbf{k}-points for Zr. The lattice constant obtained within the quasiharmonic approximation calculations of equilibrium density at $T\,{=}\,300$~K is used for QMD simulation at normal conditions. The corresponding energy obtained in QMD simulation of 250 atoms is $U_{300}\,{=}\,-813.79$~kJ/mol, while pressure is almost zero. This value of internal energy is used for $\Delta H$ calculation  along the zero isobar.
%The standard error is estimated as 0.1~GPa for pressure.
%, and ???~kJ/mol for internal energy. 
%As can be seen, the convergence of energy is achieved, and the pressure is close to zero. 
%To calculate $\Delta H$ along the zero isobar, we used the value $U_{300} = -813.79$~kJ/mol. In the solid phase, the tests show that simulations with 128 atoms and the Baldereschi's mean-value \textbf{k}-point sampling~\cite{Baldereschi:PRB:1973} and 250 atoms in the supercell and the $2 \times 2 \times 2$ Monkhorst--Pack grid~\cite{Monkhorst:PRB:1976} for the Brillouin zone sampling show the pressure and energy values included in the error. 
%In the liquid phase, tests show that use of thes Baldereschi's mean-value point also provides good accuracy contain for liquid molybdenum.

Table~\ref{tab:conv-sigma} contains convergence tests for calculated dc conductivity on system size and \textbf{k}-point sampling.

\begin{table*}[t!]
\caption{\label{tab:conv} Representative convergence tests for the calculated EOS of Zr with respect to the \textbf{k}-point grids and the number of atoms $N$. The symbols M and B indicate whether the Monkhorst--Pack scheme of the \textbf{k}-point grid or the Baldereschi mean-value \textbf{k}-point $\{ 1/4, 1/4, 1/4 \}$ is used.
}
%\begin{ruledtabular}
\begin{tabular}{cccccc}
\hline
$\rho$~(g/cm$^3$) & $T$~(kK) & $N$ & $\textbf{k}$-point & $P$~(GPa) & $U$~(kJ/mol)\\
\hline
6.455 & 0.3 & 128 & M~$[2 \times 2 \times 2]$ & 0.07 & -814.34\\
6.455 & 0.3 & 250 & B~$\{ 1/4, 1/4, 1/4 \}$ & 0.01 & -813.79\\
%\hline
6.500 & 0.3 & 128 & M~$[2 \times 2 \times 2]$ & 0.54 & -814.46\\
6.500 & 0.3 & 250 & B~$\{ 1/4, 1/4, 1/4 \}$ & 0.62 & -813.70\\
\hline
6.450 & 1.0 & 128 & M~$[2 \times 2 \times 2]$ & 0.94 & -793.82\\
6.450 & 1.0 & 250 & B~$\{ 1/4, 1/4, 1/4 \}$ & 1.30 & -793.62\\
%\hline
6.500 & 1.0 & 128 & M~$[2 \times 2 \times 2]$ & 1.75 & -792.72\\
6.500 & 1.0 & 250 & B~$\{ 1/4, 1/4, 1/4 \}$ & 1.97 & -793.39\\
\hline
6.500 & 1.25 & 250 & B~$\{ 1/4, 1/4, 1/4 \}$ & 0.09 & -783.59\\
6.500 & 1.25 & 250 & M~$[2 \times 2 \times 2]$ & 0.09 & -783.66\\
\hline
6.300 & 1.5 & 128 & M~$[2 \times 2 \times 2]$ & 0.10 & -776.55\\
6.300 & 1.5 & 250 & B~$\{ 1/4, 1/4, 1/4 \}$ & 0.12 & -776.46\\
6.300 & 1.5 & 432 & B~$\{ 1/4, 1/4, 1/4 \}$ & 0.08 & -776.67\\
%\hline
6.400 & 1.5 & 128 & M~$[2 \times 2 \times 2]$ & 1.32 & -777.17\\
6.400 & 1.5 & 250 & B~$\{ 1/4, 1/4, 1/4 \}$ & 1.33 & -777.22\\
6.400 & 1.5 & 432 & B~$\{ 1/4, 1/4, 1/4 \}$ & 1.30 & -777.24\\
\hline
6.400 & 2.0 & 128 & M~$[2 \times 2 \times 2]$ & 2.69 & -762.24\\
6.400 & 2.0 & 250 & B~$\{ 1/4, 1/4, 1/4 \}$ & 2.62 & -762.54\\
\hline
\end{tabular}
%\end{ruledtabular}
\end{table*}

\begin{table*}[t!]
\caption{\label{tab:conv-sigma} Representative convergence tests for the calculated dc conductivity of Zr in the solid and liquid phase with respect to the \textbf{k}-point grids and the number of atoms $N$.
}
\begin{tabular}{ccccc}
\hline
$\rho$~(g/cm$^3$) & $T$~(kK) & $N$ & $\textbf{k}$-point & $\sigma_{dc}$~(10$^3$ Sm/m)\\
%\hline
%6.3 & 1.50 & 250 & M~$[2 \times 2 \times 2]$ & 907.68\\
%6.3 & 1.50 & 250 & M~$[3 \times 3 \times 3]$ & 906.5\\
%6.3 & 1.50 & 432 & B~$\{ 1/4, 1/4, 1/4 \}$ & 905.21\\
\hline
6.4 & 1.50 & 128 & M~$[2 \times 2 \times 2]$ & 981.84\\
6.4 & 1.50 & 250 & B~$\{ 1/4, 1/4, 1/4 \}$ & 973.05\\
6.4 & 1.50 & 250 & M~$[2 \times 2 \times 2]$ & 952.82\\
6.4 & 1.50 & 250 & M~$[3 \times 3 \times 3]$ & 953.03\\
6.4 & 1.50 & 432 & M~$[2 \times 2 \times 2]$ & 952.31\\
\hline
6.0 & 2.25 & 128 & M~$[2 \times 2 \times 2]$ & 756.98\\
6.0 & 2.25 & 250 & B~$\{ 1/4, 1/4, 1/4 \}$ & 757.08\\
6.0 & 2.25 & 250 & M~$[2 \times 2 \times 2]$ & 759.61\\
6.0 & 2.25 & 250 & M~$[3 \times 3 \times 3]$ & 757.53\\
\hline
\end{tabular}
\end{table*}

\begin{table}[h]
\caption{\label{tab:eos} Equation of state data and dc conductivity for Zr from QMD calculations.}
%\begin{ruledtabular}
\begin{tabular}{ccccc}
$\rho$~(g/cm$^3$) & $T$~(kK) & $P$~(GPa) & $U$~(kJ/mol) & $\sigma_{dc}$~(10$^3$ Sm/m)\\
\hline
\multicolumn{5}{c}{$\alpha$ phase} \\    
\hline 

6.40 & 0.75 & 0.08 & -801.18 & -\\
6.50 & 0.75 & 1.40 & -801.23 & -\\
6.60 & 0.75 & 2.75 & -800.62 & -\\

6.40 & 1.00 & 0.64 & -793.59 & -\\
6.45 & 1.00 & 1.30 & -793.61 & -\\
6.50 & 1.00 & 1.97 & -793.39 & -\\

6.35 & 1.15 & 0.29 & -788.63 & -\\
6.40 & 1.15 & 0.98 & -788.95 & -\\
6.45 & 1.15 & 1.65 & -788.96 & -\\

\hline
\multicolumn{5}{c}{$\beta$ phase} \\    
\hline

6.35 & 1.25 & 0.09 & -783.59 & 985.61\\
6.42 & 1.25 & 0.96 & -784.19 & 986.02\\
6.50 & 1.25 & 1.99 & -784.21 & 1006.90\\
6.70 & 1.25 & 4.66 & -784.20 & 1081.86\\

6.30 & 1.50 & 0.12 & -776.46 & 915.24\\
6.35 & 1.50 & 0.72 & -776.77 & 946.55\\
6.40 & 1.50 & 1.36 & -777.22 & 952.82\\
6.50 & 1.50 & 2.60 & -777.51 & 977.42\\
6.70 & 1.50 & 5.27 & -777.25 & 1031.11\\

6.25 & 1.75 & 0.17 & -768.80 & 883.45\\
6.35 & 1.75 & 1.34 & -769.64 & 897.75\\
6.45 & 1.75 & 2.57 & -770.20 & 926.24\\
6.55 & 1.75 & 3.88 & -770.33 & 952.68\\
6.75 & 1.75 & 6.63 & -769.94 & 1003.53\\

6.40 & 2.00 & 2.62 & -762.54 & 886.16\\
6.60 & 2.00 & 5.18 & -763.03 & 929.34\\
6.80 & 2.00 & 7.98 & -762.58 & 980.83\\

\hline
\multicolumn{5}{c}{Liquid} \\ 
\hline

6.00 & 2.15 & 0.01 & -741.29 & -\\
6.00 & 2.25 & 0.26 & -738.63 & 759.61\\
6.00 & 2.35 & 0.59 & -734.79 & 761.81\\
6.00 & 3.20 & 2.90 & -706.10 & 751.69\\
6.00 & 4.00 & 5.19 & -678.31 & 742.53\\

5.75 & 3.00 & 0.02 & -708.42 & 722.48\\
5.75 & 3.10 & 0.31 & -705.03 & 725.65\\
5.75 & 3.20 & 0.53 & -702.11 & 723.62\\
5.75 & 4.00 & 2.74 & -674.85 & 715.82\\
5.75 & 5.00 & 5.37 & -640.68 & 707.81\\

5.50 & 3.80 & 0.05 & -675.01 & 688.88\\
5.50 & 3.90 & 0.50 & -672.34 & -\\
5.50 & 4.00 & 0.63 & -668.95 & 689.30\\
5.50 & 5.00 & 3.29 & -633.78 & 679.56\\
5.50 & 6.00 & 5.65 & -599.25 & 666.72\\

\end{tabular}
%\end{ruledtabular}
\end{table}

\appendix
\section{\label{app:eos}Equation of state}

The EOS data derived from the QMD simulations are presented in Table~\ref{tab:eos}, as well as calculated dc conductivity along the isotherms for $\beta$-Zr and along the isochors for liquid metal. 

%\bibliographystyle{aipnum4-1}
\bibliographystyle{elsarticle-num}
%\bibliography{Zr_melting}
%\bibliography{zr_melting,vaspref,paramonov}

\begin{thebibliography}{100}
\expandafter\ifx\csname url\endcsname\relax
  \def\url#1{\texttt{#1}}\fi
\expandafter\ifx\csname urlprefix\endcsname\relax\def\urlprefix{URL }\fi
\expandafter\ifx\csname href\endcsname\relax
  \def\href#1#2{#2} \def\path#1{#1}\fi

\bibitem{tonkov2018phase}
E.~Y. Tonkov, E.~Ponyatovsky, Phase transformations of elements under high
  pressure, CRC press, 2018.

\bibitem{sikka1982omega}
S.~Sikka, Y.~Vohra, R.~Chidambaram, Omega phase in materials, Progress in
  Materials Science 27~(3-4) (1982) 245--310.

\bibitem{dewaele2015high}
A.~Dewaele, V.~Stutzmann, J.~Bouchet, F.~Bottin, F.~Occelli, M.~Mezouar, High
  pressure-temperature phase diagram and equation of state of titanium,
  Physical Review B 91~(13) (2015) 134108.

\bibitem{Anzellini:PRB:2020}
S.~Anzellini, F.~Bottin, J.~Bouchet, A.~Dewaele,
  \href{https://link.aps.org/doi/10.1103/PhysRevB.102.184105}{Phase transitions
  and equation of state of zirconium under high pressure}, Phys. Rev. B 102
  (2020) 184105.
\newblock \href {https://doi.org/10.1103/PhysRevB.102.184105}
  {\path{doi:10.1103/PhysRevB.102.184105}}.

\bibitem{parisiades2019melting}
P.~Parisiades, F.~Cova, G.~Garbarino, Melting curve of elemental zirconium,
  Physical Review B 100~(5) (2019) 054102.

\bibitem{hellman2011lattice}
O.~Hellman, I.~Abrikosov, S.~Simak, Lattice dynamics of anharmonic solids from
  first principles, Physical Review B 84~(18) (2011) 180301.

\bibitem{Bonnell:RU:1972}
D.~W. Bonnell, Property measurements at high temperatures--levitation
  calorimetry studies of liquid metals, Ph.D. Thesis (1972) 1--72.

\bibitem{hixson1993thermophysical}
R.~Hixson, M.~Winkler, Thermophysical properties of liquid platinum,
  International journal of thermophysics 14~(3) (1993) 409--416.

\bibitem{boivineau1996thermophysical}
M.~Boivineau, H.~Colin, J.~Vermeulen, T.~Th{\'e}venin, Thermophysical
  properties of solid and liquid thorium, International journal of
  thermophysics 17~(5) (1996) 1001--1010.

\bibitem{obendrauf1993measurements}
W.~Obendrauf, E.~Kaschnitz, G.~Pottlacher, H.~J{\"a}ger, Measurements of
  thermophysical properties of nickel with a new highly sensitive pyrometer,
  International journal of thermophysics 14~(3) (1993) 417--426.

\bibitem{seifter2001microsecond}
A.~Seifter, F.~Sachsenhofer, S.~Krishnan, G.~Pottlacher, Microsecond laser
  polarimetry for emissivity measurements on liquid metals at high
  temperatures---application to niobium, International journal of thermophysics
  22~(5) (2001) 1537--1547.

\bibitem{Gathers:RPP:1986}
G.~R. Gathers, \href{http://stacks.iop.org/0034-4885/49/i=4/a=001}{Dynamic
  methods for investigating thermophysical properties of matter at very high
  temperatures and pressures}, Reports on Progress in Physics 49~(4) (1986)
  341.

\bibitem{Boivineau:IJMPT:2006}
M.~Boivineau, G.~Pottlacher,
  \href{https://www.inderscienceonline.com/doi/abs/10.1504/IJMPT.2006.009468}{Thermophysical
  properties of metals at very high temperatures obtained by dynamic heating
  techniques: recent advances}, International Journal of Materials and Product
  Technology 26~(3-4) (2006) 217--246.
\newblock \href {https://doi.org/10.1504/IJMPT.2006.009468}
  {\path{doi:10.1504/IJMPT.2006.009468}}.

\bibitem{rhim1993electrostatic}
W.-K. Rhim, S.~K. Chung, D.~Barber, K.~F. Man, G.~Gutt, A.~Rulison, R.~E.
  Spjut, An electrostatic levitator for high-temperature containerless
  materials processing in 1-g, Review of Scientific Instruments 64~(10) (1993)
  2961--2970.

\bibitem{ishikawa2005non}
T.~Ishikawa, P.-F. Paradis, T.~Itami, S.~Yoda, Non-contact thermophysical
  property measurements of refractory metals using an electrostatic levitator,
  Measurement Science and Technology 16~(2) (2005) 443.

\bibitem{bradshaw2005machine}
R.~Bradshaw, D.~Schmidt, J.~Rogers, K.~Kelton, R.~Hyers, Machine vision for
  high-precision volume measurement applied to levitated containerless material
  processing, Review of scientific instruments 76~(12) (2005) 125108.

\bibitem{mauro2011highly}
N.~Mauro, K.~Kelton, A highly modular beamline electrostatic levitation
  facility, optimized for in situ high-energy x-ray scattering studies of
  equilibrium and supercooled liquids, Review of Scientific Instruments 82~(3)
  (2011) 035114.

\bibitem{lee2013crystal}
G.~W. Lee, S.~Jeon, D.-H. Kang, Crystal--liquid interfacial free energy of
  supercooled liquid fe using a containerless technique, Crystal growth \&
  design 13~(4) (2013) 1786--1792.

\bibitem{paradis2014materials}
P.-F. Paradis, T.~Ishikawa, G.-W. Lee, D.~Holland-Moritz, J.~Brillo, W.-K.
  Rhim, J.~T. Okada, Materials properties measurements and particle beam
  interactions studies using electrostatic levitation, Materials Science and
  Engineering: R: Reports 76 (2014) 1--53.

\bibitem{millot2002high}
F.~Millot, J.-C. Rifflet, V.~Sarou-Kanian, G.~Wille, High-temperature
  properties of liquid boron from contactless techniques, International journal
  of Thermophysics 23~(5) (2002) 1185--1195.

\bibitem{wille2002thermophysical}
G.~Wille, F.~Millot, J.~Rifflet, Thermophysical properties of containerless
  liquid iron up to 2500 k, International Journal of Thermophysics 23~(5)
  (2002) 1197--1206.

\bibitem{egry2000thermophysical}
I.~Egry, Thermophysical property measurements in microgravity, High
  Temperatures High Pressures(UK) 32~(2) (2000) 127--134.

\bibitem{lee2013crystal_Ti}
G.~W. Lee, S.~Jeon, C.~Park, D.-H. Kang, Crystal--liquid interfacial free
  energy and thermophysical properties of pure liquid ti using electrostatic
  levitation: Hypercooling limit, specific heat, total hemispherical
  emissivity, density, and interfacial free energy, The Journal of Chemical
  Thermodynamics 63 (2013) 1--6.

\bibitem{ishikawa2001new}
T.~Ishikawa, P.-F. Paradis, S.~Yoda, New sample levitation initiation and
  imaging techniques for the processing of refractory metals with an
  electrostatic levitator furnace, Review of Scientific Instruments 72~(5)
  (2001) 2490--2495.

\bibitem{ishikawa2003thermophysical}
T.~Ishikawa, P.-F. Paradis, T.~Itami, S.~Yoda, Thermophysical properties of
  liquid refractory metals: comparison between hard sphere model calculation
  and electrostatic levitation measurements, The Journal of chemical physics
  118~(17) (2003) 7912--7920.

\bibitem{Wimmer:JPCM:2010}
E.~Wimmer, R.~Najafabadi, G.~A.~Y. Jr, J.~D. Ballard, T.~M. Angeliu,
  J.~Vollmer, J.~J. Chambers, H.~Niimi, J.~B. Shaw, C.~Freeman, M.~Christensen,
  W.~Wolf, P.~Saxe, \href{https://doi.org/10.1088/0953-8984/22/38/384215}{Ab
  initio calculations for industrial materials engineering: successes and
  challenges}, Journal of Physics: Condensed Matter 22~(38) (2010) 384215.
\newblock \href {https://doi.org/10.1088/0953-8984/22/38/384215}
  {\path{doi:10.1088/0953-8984/22/38/384215}}.

\bibitem{Tu:NME:2018}
R.~Tu, Q.~Liu, C.~Zeng, Y.~Li, W.~Xiao,
  \href{https://www.sciencedirect.com/science/article/pii/S2352179118300413}{First
  principles study of point defect effects on iodine diffusion in zirconium},
  Nuclear Materials and Energy 16 (2018) 238--244.
\newblock \href {https://doi.org/https://doi.org/10.1016/j.nme.2018.07.006}
  {\path{doi:https://doi.org/10.1016/j.nme.2018.07.006}}.

\bibitem{Xin:JNM:2009}
X.~Xin, W.~Lai, B.~Liu,
  \href{https://www.sciencedirect.com/science/article/pii/S0022311509006485}{Point
  defect properties in hcp and bcc zr with trace solute nb revealed by ab
  initio calculations}, Journal of Nuclear Materials 393~(1) (2009) 197--202.
\newblock \href {https://doi.org/https://doi.org/10.1016/j.jnucmat.2009.06.005}
  {\path{doi:https://doi.org/10.1016/j.jnucmat.2009.06.005}}.

\bibitem{Xiong:JNM:2014}
W.~Xiong, W.~Xie, D.~Morgan,
  \href{https://www.sciencedirect.com/science/article/pii/S0022311514003808}{Thermodynamic
  evaluation of the np--zr system using calphad and ab initio methods}, Journal
  of Nuclear Materials 452~(1) (2014) 569--577.
\newblock \href {https://doi.org/https://doi.org/10.1016/j.jnucmat.2014.06.023}
  {\path{doi:https://doi.org/10.1016/j.jnucmat.2014.06.023}}.

\bibitem{Khanal:NME:2021}
R.~Khanal, N.~Ayers, N.~Jerred, M.~T. Benson, R.~D. Mariani, I.~Charit,
  S.~Choudhury,
  \href{https://www.sciencedirect.com/science/article/pii/S2352179121000107}{Role
  of zirconium in neodymium-dopants interactions within uranium-based metallic
  fuels}, Nuclear Materials and Energy 26 (2021) 100912.
\newblock \href {https://doi.org/https://doi.org/10.1016/j.nme.2021.100912}
  {\path{doi:https://doi.org/10.1016/j.nme.2021.100912}}.

\bibitem{Christensen:JNM:2015}
M.~Christensen, W.~Wolf, C.~Freeman, E.~Wimmer, R.~Adamson, L.~Hallstadius,
  P.~Cantonwine, E.~Mader,
  \href{https://www.sciencedirect.com/science/article/pii/S0022311515001063}{Diffusion
  of point defects, nucleation of dislocation loops, and effect of hydrogen in
  hcp-zr: Ab initio and classical simulations}, Journal of Nuclear Materials
  460 (2015) 82--96.
\newblock \href {https://doi.org/https://doi.org/10.1016/j.jnucmat.2015.02.013}
  {\path{doi:https://doi.org/10.1016/j.jnucmat.2015.02.013}}.

\bibitem{Peng:JNM:2012}
Q.~Peng, W.~Ji, H.~Huang, S.~De,
  \href{https://www.sciencedirect.com/science/article/pii/S0022311512002899}{Stability
  of self-interstitial atoms in hcp-zr}, Journal of Nuclear Materials 429~(1)
  (2012) 233--236.
\newblock \href {https://doi.org/https://doi.org/10.1016/j.jnucmat.2012.06.010}
  {\path{doi:https://doi.org/10.1016/j.jnucmat.2012.06.010}}.

\bibitem{Willaime:JNM:2003}
F.~Willaime,
  \href{https://www.sciencedirect.com/science/article/pii/S0022311503003647}{Ab
  initio study of self-interstitials in hcp-zr}, Journal of Nuclear Materials
  323~(2) (2003) 205--212, proceedings of the Second IEA Fusion Materials
  Agreement Workshop on Modeling and Experimental Validation.
\newblock \href {https://doi.org/https://doi.org/10.1016/j.jnucmat.2003.08.005}
  {\path{doi:https://doi.org/10.1016/j.jnucmat.2003.08.005}}.

\bibitem{Hao:PRB:2008}
Y.-J. Hao, L.~Zhang, X.-R. Chen, L.-C. Cai, Q.~Wu, D.~Alf\`e,
  \href{https://link.aps.org/doi/10.1103/PhysRevB.78.134101}{Ab initio
  calculations of the thermodynamics and phase diagram of zirconium}, Phys.
  Rev. B 78 (2008) 134101.
\newblock \href {https://doi.org/10.1103/PhysRevB.78.134101}
  {\path{doi:10.1103/PhysRevB.78.134101}}.

\bibitem{Zhou:JNM:2018}
W.~Zhou, J.~Tian, Q.~Feng, J.~Zheng, X.~Liu, J.~Xue, D.~Qian, S.~Peng,
  \href{https://www.sciencedirect.com/science/article/pii/S0022311518303210}{Molecular
  dynamics simulations of high-energy displacement cascades in hcp-zr}, Journal
  of Nuclear Materials 508 (2018) 540--545.
\newblock \href {https://doi.org/https://doi.org/10.1016/j.jnucmat.2018.06.002}
  {\path{doi:https://doi.org/10.1016/j.jnucmat.2018.06.002}}.

\bibitem{Li:NME:2019}
Y.~Li, H.~Chen, Y.~Chen, Y.~Wang, L.~Shao, W.~Xiao,
  \href{https://www.sciencedirect.com/science/article/pii/S2352179119300158}{Point
  defect effects on tensile strength of $\alpha$-zirconium studied by molecular
  dynamics simulations}, Nuclear Materials and Energy 20 (2019) 100683.
\newblock \href {https://doi.org/https://doi.org/10.1016/j.nme.2019.100683}
  {\path{doi:https://doi.org/10.1016/j.nme.2019.100683}}.

\bibitem{Moore:JNM:2015}
A.~Moore, B.~Beeler, C.~Deo, M.~Baskes, M.~Okuniewski,
  \href{https://www.sciencedirect.com/science/article/pii/S0022311515302683}{Atomistic
  modeling of high temperature uranium--zirconium alloy structure and
  thermodynamics}, Journal of Nuclear Materials 467 (2015) 802--819.
\newblock \href {https://doi.org/https://doi.org/10.1016/j.jnucmat.2015.10.016}
  {\path{doi:https://doi.org/10.1016/j.jnucmat.2015.10.016}}.

\bibitem{Fang:CMS:2008}
H.~Fang, X.~Hui, G.~Chen, R.~{\"O}ttking, Y.~Liu, J.~Schaefer, Z.~Liu,
  \href{https://www.sciencedirect.com/science/article/pii/S092702560800147X}{Ab
  initio molecular dynamics simulation for structural transition of zr during
  rapid quenching processes}, Computational Materials Science 43~(4) (2008)
  1123--1129.
\newblock \href
  {https://doi.org/https://doi.org/10.1016/j.commatsci.2008.03.011}
  {\path{doi:https://doi.org/10.1016/j.commatsci.2008.03.011}}.

\bibitem{jakse2007short}
N.~Jakse, O.~Le~Bacq, A.~Pasturel, Short-range order of liquid and undercooled
  metals: Ab initio molecular dynamics study, Journal of non-crystalline solids
  353~(32-40) (2007) 3684--3688.

\bibitem{Jakse:JNS:2007}
N.~Jakse, O.~L. Bacq, A.~Pasturel, Short-range order of liquid and undercooled
  metals: Ab initio molecular dynamics study, Journal of Non-Crystalline Solids
  353~(32) (2007) 3684--3688.
\newblock \href
  {https://doi.org/https://doi.org/10.1016/j.jnoncrysol.2007.05.131}
  {\path{doi:https://doi.org/10.1016/j.jnoncrysol.2007.05.131}}.

\bibitem{Minakov:AIPADV:2018}
D.~V. Minakov, M.~A. Paramonov, P.~R. Levashov,
  \href{https://doi.org/10.1063/1.5062152}{Ab initio inspection of
  thermophysical experiments for molybdenum near melting}, AIP Advances 8~(12)
  (2018) 125012.
\newblock \href
  {https://doi.org/10.1063/1.5062152} {\path{doi:10.1063/1.5062152}}.

\bibitem{Minakov:PRB:2018}
D.~V. Minakov, M.~A. Paramonov, P.~R. Levashov,
  \href{https://link.aps.org/doi/10.1103/PhysRevB.97.024205}{Consistent
  interpretation of experimental data for expanded liquid tungsten near the
  liquid-gas coexistence curve}, Phys. Rev. B 97 (2018) 024205.
\newblock \href {https://doi.org/10.1103/PhysRevB.97.024205}
  {\path{doi:10.1103/PhysRevB.97.024205}}.

\bibitem{Minakov:HTHP:2020}
D.~V. Minakov, M.~A. Paramonov, P.~R. Levashov, Interpretation of pulse-heating
  experiments for rhenium by quantum molecular dynamics., High
  Temperatures--High Pressures 49~(1-2) (2020) 211--219.

\bibitem{Clerouin:PRE:2008}
J.~Cl\'erouin, P.~Renaudin, P.~Noiret,
  \href{https://link.aps.org/doi/10.1103/PhysRevE.77.026409}{Experiments and
  simulations on hot expanded boron}, Phys. Rev. E 77 (2008) 026409.
\newblock \href {https://doi.org/10.1103/PhysRevE.77.026409}
  {\path{doi:10.1103/PhysRevE.77.026409}}.

\bibitem{Miljacic:CALPH:2015}
L.~Miljacic, S.~Demers, Q.-J. Hong, A.~van~de Walle,
  \href{http://www.sciencedirect.com/science/article/pii/S0364591615300134}{Equation
  of state of solid, liquid and gaseous tantalum from first principles},
  Calphad 51~(Supplement C) (2015) 133 -- 143.
\newblock \href {https://doi.org/https://doi.org/10.1016/j.calphad.2015.08.005}
  {\path{doi:https://doi.org/10.1016/j.calphad.2015.08.005}}.

\bibitem{Clerouin:PRB:2008}
J.~Cl\'erouin, P.~Noiret, V.~N. Korobenko, A.~D. Rakhel,
  \href{https://link.aps.org/doi/10.1103/PhysRevB.78.224203}{Direct
  measurements and ab initio simulations for expanded fluid aluminum in the
  metal-nonmetal transition range}, Phys. Rev. B 78 (2008) 224203.
\newblock \href {https://doi.org/10.1103/PhysRevB.78.224203}
  {\path{doi:10.1103/PhysRevB.78.224203}}.

\bibitem{Mermin:PR:1965}
N.~D. Mermin, \href{https://link.aps.org/doi/10.1103/PhysRev.137.A1441}{Thermal
  properties of the inhomogeneous electron gas}, Phys. Rev. 137 (1965)
  A1441--A1443.
\newblock \href {https://doi.org/10.1103/PhysRev.137.A1441}
  {\path{doi:10.1103/PhysRev.137.A1441}}.

\bibitem{Kresse:PRB:1993}
G.~Kresse, J.~Hafner, Ab initio molecular dynamics for liquid metals, Phys.
  Rev. B 47 (1993) 558.

\bibitem{Kresse:PRB:1994}
G.~Kresse, J.~Hafner, Ab initio molecular-dynamics simulation of the
  liquid-metal-amorphous-semiconductor transition in germanium, Phys. Rev. B 49
  (1994) 14251.

\bibitem{Kresse:PRB:1996}
G.~Kresse, J.~Furthm\"uller,
  \href{http://link.aps.org/doi/10.1103/PhysRevB.54.11169}{Efficient iterative
  schemes for \textit{ab initio} total-energy calculations using a plane-wave
  basis set}, Phys. Rev. B 54 (1996) 11169--11186.
\newblock \href {https://doi.org/10.1103/PhysRevB.54.11169}
  {\path{doi:10.1103/PhysRevB.54.11169}}.

\bibitem{Kresse:CMS:1996}
G.~Kresse, J.~Furthm{\"u}ller,
  \href{http://www.sciencedirect.com/science/article/pii/0927025696000080}{Efficiency
  of ab-initio total energy calculations for metals and semiconductors using a
  plane-wave basis set}, Computational Materials Science 6~(1) (1996) 15 -- 50.
\newblock \href {https://doi.org/10.1016/0927-0256(96)00008-0}
  {\path{doi:10.1016/0927-0256(96)00008-0}}.

\bibitem{Blochl:PR:1994}
P.~E. Bl\"ochl,
  \href{http://link.aps.org/doi/10.1103/PhysRevB.50.17953}{Projector
  augmented-wave method}, Phys. Rev. B 50 (1994) 17953--17979.
\newblock \href {https://doi.org/10.1103/PhysRevB.50.17953}
  {\path{doi:10.1103/PhysRevB.50.17953}}.

\bibitem{Perdew:PRL:1996}
J.~P. Perdew, K.~Burke, M.~Ernzerhof, Generalized gradient approximation made
  simple, Phys. Rev. Lett. 77 (1996) 3865.

\bibitem{Perdew:PRL:1997}
J.~P. Perdew, K.~Burke, M.~Ernzerhof, Erratum: Generalized gradient
  approximation made simple, Phys. Rev. Lett. 78 (1997) 1396.

\bibitem{Nose:JCP:1984}
S.~Nos{\'e}, \href{http://dx.doi.org/10.1063/1.447334}{A unified formulation of
  the constant temperature molecular dynamics methods}, The Journal of Chemical
  Physics 81~(1) (1984) 511--519.
\newblock \href {https://doi.org/10.1063/1.447334}
  {\path{doi:10.1063/1.447334}}.

\bibitem{CRC:2005}
D.~R. Lide (Ed.), CRC Handbook of Chemistry and Physics, Internet Version 2005,
  CRC Press, Boca Raton, FL, 2005.

\bibitem{Baldereschi:PRB:1973}
A.~Baldereschi,
  \href{https://link.aps.org/doi/10.1103/PhysRevB.7.5212}{Mean-value point in
  the brillouin zone}, Phys. Rev. B 7 (1973) 5212--5215.
\newblock \href {https://doi.org/10.1103/PhysRevB.7.5212}
  {\path{doi:10.1103/PhysRevB.7.5212}}.

\bibitem{Armiento:PRB:2005}
R.~Armiento, A.~E. Mattsson,
  \href{https://link.aps.org/doi/10.1103/PhysRevB.72.085108}{Functional
  designed to include surface effects in self-consistent density functional
  theory}, Phys. Rev. B 72 (2005) 085108.
\newblock \href {https://doi.org/10.1103/PhysRevB.72.085108}
  {\path{doi:10.1103/PhysRevB.72.085108}}.

\bibitem{Ceperley:PRL:1980}
D.~M. Ceperley, B.~J. Alder,
  \href{https://link.aps.org/doi/10.1103/PhysRevLett.45.566}{Ground state of
  the electron gas by a stochastic method}, Phys. Rev. Lett. 45 (1980)
  566--569.
\newblock \href {https://doi.org/10.1103/PhysRevLett.45.566}
  {\path{doi:10.1103/PhysRevLett.45.566}}.

\bibitem{Knyazev_2018}
D.~V. Knyazev, P.~R. Levashov, Thermodynamic, transport, and optical properties
  of dense silver plasma calculated using the {GreeKuP} code, Contrib. Plasma
  Phys. 59~(3) (2018) 345--353.
\newblock \href {https://doi.org/10.1002/ctpp.201800084}
  {\path{doi:10.1002/ctpp.201800084}}.

\bibitem{Desjarlais:PRE:2002}
M.~P. Desjarlais, J.~D. Kress, L.~A. Collins, Electrical conductivity for warm,
  dense aluminum plasmas and liquids, Phys. Rev. E 66~(2) (2002) 025401.
\newblock \href {https://doi.org/10.1103/physreve.66.025401}
  {\path{doi:10.1103/physreve.66.025401}}.

\bibitem{Knyazev2013}
D.~V. Knyazev, P.~R. Levashov, Ab initio calculation of transport and optical
  properties of aluminum: Influence of simulation parameters, Comput. Mater.
  Sci. 79 (2013) 817--829.
\newblock \href {https://doi.org/10.1016/j.commatsci.2013.04.066}
  {\path{doi:10.1016/j.commatsci.2013.04.066}}.

\bibitem{Kowalski2007}
P.~M. Kowalski, S.~Mazevet, D.~Saumon, M.~Challacombe, Equation of state and
  optical properties of warm dense helium, Phys. Rev. B 76~(7) (2007) 075112.
\newblock \href {https://doi.org/10.1103/physrevb.76.075112}
  {\path{doi:10.1103/physrevb.76.075112}}.

\bibitem{Monkhorst:PRB:1976}
H.~J. Monkhorst, J.~D. Pack,
  \href{https://link.aps.org/doi/10.1103/PhysRevB.13.5188}{Special points for
  brillouin-zone integrations}, Phys. Rev. B 13 (1976) 5188--5192.
\newblock \href {https://doi.org/10.1103/PhysRevB.13.5188}
  {\path{doi:10.1103/PhysRevB.13.5188}}.

\bibitem{Toll_1956}
J.~S. Toll, Causality and the dispersion relation: Logical foundations, Phys.
  Rev. 104~(6) (1956) 1760--1770.
\newblock \href {https://doi.org/10.1103/physrev.104.1760}
  {\path{doi:10.1103/physrev.104.1760}}.

\bibitem{peterson1958surface}
A.~Peterson, H.~Kedesdy, P.~Keck, E.~Schwarz, Surface tension of titanium,
  zirconium, and hafnium, Journal of Applied Physics 29~(2) (1958) 213--216.

\bibitem{allen1963surface}
B.~Allen, The surface tension of liquid transition metals at their melting
  points, Trans. AIME 227 (1963).

\bibitem{yelyutin1964viscosity}
V.~Yelyutin, M.~Maurakh, I.~Penkov, Viscosity of liquid zirconium (1964).

\bibitem{steinberg1974simple}
D.~Steinberg, A simple relationship between the temperature dependence of the
  density of liquid metals and their boiling temperatures, Metallurgical
  Transactions 5~(6) (1974) 1341--1343.

\bibitem{guillermet1987critical}
A.~F. Guillermet, Critical evaluation of the thermodynamic properties of
  zirconium, High Temperatures-High Pressures 19~(2) (1987) 119--160.

\bibitem{paradis1999thermophysical}
P.-F. Paradis, W.-K. ~, Thermophysical properties of zirconium at high
  temperature, Journal of materials research 14~(9) (1999) 3713--3719.

\bibitem{yoo2015uncertainty}
H.~Yoo, C.~Park, S.~Jeon, S.~Lee, G.~W. Lee, Uncertainty evaluation for density
  measurements of molten ni, zr, nb and hf by using a containerless method,
  Metrologia 52~(5) (2015) 677.

\bibitem{Ishikawa:RSI:2001}
T.~Ishikawa, P.-F. Paradis, S.~Yoda, New sample levitation initiation and
  imaging techniques for the processing of refractory metals with an
  electrostatic levitator furnace, Review of Scientific Instruments 72~(5)
  (2001) 2490--2495.
\newblock \href {https://doi.org/10.1063/1.1368861}
  {\path{doi:10.1063/1.1368861}}.

\bibitem{korobenko2001temperature}
V.~N. Korobenko, A.~I. Savvatimskii, Temperature dependence of the density and
  electrical resistivity of liquid zirconium up to 4100 k, High Temperature
  39~(4) (2001) 525--531.

\bibitem{chase1998nist}
M.~Chase~Jr, Nist-janaf thermochemical tables, J. Phys. Chem. Ref. Dara
  Monogrph 9 (1998) 1019.

\bibitem{arblaster2013thermodynamic}
J.~Arblaster, Thermodynamic properties of zirconium, Calphad 43 (2013) 32--39.

\bibitem{Gurvich:IVTANTHERMO:1982}
L.~V. Gurvich, I.~V. Veits, V.~A. Medvedev, Thermodynamic Properties of
  Individual Substances. Volume 4 [in Russian], Nauka, Moscow, 1982.

\bibitem{korobenko2001experimental}
V.~N. Korobenko, A.~I. Savvatimski, K.~K. Sevostyanov, Experimental
  investigation of solid and liquid zirconium, High Temperatures -- High
  Pressures 33~(6) (2001) 647--658.

\bibitem{brunner2003normal}
C.~Brunner, C.~Cagran, A.~Seifter, G.~Pottlacher, The normal spectral
  emissivity at a wavelength of 684.5 nm and thermophysical properties of
  liquid zirconium up to the end of the stable liquid phase, in: AIP Conference
  Proceedings, Vol. 684, American Institute of Physics, 2003, pp. 771--776.

\bibitem{korobenko1991properties}
V.~N. Korobenko, A.~Savvatimskij, Properties of solid and liquid zirconium,
  Teplofizika Vysokikh Temperatur 29~(5) (1991) 883--886.

\bibitem{korobenko2002zirconium}
V.~Korobenko, M.~Agranat, S.~Ashitkov, A.~Savvatimskiy, Zirconium and iron
  densities in a wide range of liquid states, International journal of
  thermophysics 23~(1) (2002) 307--318.

\bibitem{gathers1986dynamic}
G.~Gathers, Dynamic methods for investigating thermophysical properties of
  matter at very high temperatures and pressures, Reports on Progress in
  Physics 49~(4) (1986) 341.

\bibitem{Hafner:JCC:2008}
J.~Hafner,
  \href{https://onlinelibrary.wiley.com/doi/abs/10.1002/jcc.21057}{Ab-initio
  simulations of materials using vasp: Density-functional theory and beyond},
  Journal of Computational Chemistry 29~(13) (2008) 2044--2078.
\newblock \href {https://doi.org/10.1002/jcc.21057}
  {\path{doi:10.1002/jcc.21057}}.

\bibitem{Harl:PRL:2009}
J.~Harl, G.~Kresse,
  \href{https://link.aps.org/doi/10.1103/PhysRevLett.103.056401}{Accurate bulk
  properties from approximate many-body techniques}, Phys. Rev. Lett. 103
  (2009) 056401.
\newblock \href {https://doi.org/10.1103/PhysRevLett.103.056401}
  {\path{doi:10.1103/PhysRevLett.103.056401}}.

\bibitem{touloukian1975thermophysical}
Y.~S. Touloukian, R.~Kirby, R.~Taylor, P.~Desai, Thermophysical Properties Of
  Matter - The TPRC Data Series. Volume 12. Thermal Expansion Metallic Elements
  And Alloys, New York: IFI/Plenum, 1975.

\bibitem{itami1984application}
T.~Itami, M.~Shimoji, Application of simple model theories to thermodynamic
  properties of liquid transition metals, Journal of Physics F: Metal Physics
  14~(2) (1984) L15.

\bibitem{douglas1958heat}
T.~B. Douglas, A.~C. Victor, Heat content of zirconium and of five compositions
  of zirconium hydride from 0 to 900 C, J. Res. Natl. Bur. Stand 61 (1958)
  13--23.

\bibitem{cezairliyan1974simultaneous}
A.~Cezairliyan, F.~Righini, Simultaneous measurements of heat capacity,
  electrical resistivity and hemispherical total emittance by a pulse heating
  technique: zirconium, 1500 to 2100 k, Journal of research of the National
  Bureau of Standards. Section A, Physics and chemistry 78~(4) (1974) 509.

\bibitem{petrova2000investigation}
I.~I. Petrova, V.~E. Peletskii, B.~N. Samsonov, Investigation of the
  thermophysical properties of zirconium by subsecond pulsed heating technique,
  High Temperature 38~(4) (2000) 560--565.

\bibitem{rosner2001enthalpy}
M.~R{\"o}sner-Kuhn, K.~Drewes, H.~Franz, M.~G. Frohberg, Enthalpy measurements
  of the solid high-temperature $\beta$-phase of titanium and zirconium by
  levitation drop calorimetry, Journal of alloys and compounds 316~(1-2) (2001)
  175--178.

\bibitem{Belov:JPCS:2018}
G.~V. Belov, S.~A. Dyachkov, P.~R. Levashov, I.~V. Lomonosov, D.~V. Minakov,
  I.~V. Morozov, M.~A. Sineva, V.~N. Smirnov,
  \href{http://stacks.iop.org/1742-6596/946/i=1/a=012120}{The
  ivtanthermo-online database for thermodynamic properties of individual
  substances with web interface}, Journal of Physics: Conference Series 946~(1)
  (2018) 012120.

\bibitem{jaeger1934exact}
F.~Jaeger, W.~Veenstra, The exact measurement of the specific heats of solid
  substances at high temperatures. vi. the specific heats of vanadium, niobium,
  tantalum and molybdenum, Recueil des Travaux Chimiques des Pays-Bas 53~(8)
  (1934) 677--687.

\bibitem{coughlin1950high}
J.~Coughlin, E.~King, High-temperature heat contents of some
  zirconium-containing substances1, Journal of the American Chemical Society
  72~(5) (1950) 2262--2265.

\bibitem{skinner1951vapor}
G.~B. Skinner, J.~W. Edwards, H.~L. Johnston, The vapor pressure of inorganic
  substances. v. zirconium between 1949 and 2054 K., Journal of the
  American Chemical Society 73~(1) (1951) 174--176.

\bibitem{adenstedt1952physical}
H.~Adenstedt, Physical, thermal and electrical properties of hafnium and high
  purity zirconium, Transactions of the American Society for Metals 44 (1952)
  949--973.

\bibitem{redmond1952enthalpies}
R.~Redmond, J.~Lones, Enthalpies and heat capacities of stainless steel (316),
  zirconium, and lithium at elevated temperatures, Tech. rep., Oak Ridge
  National Lab. (1952).

\bibitem{fieldhouse1961armour}
I.~Fieldhouse, J.~Lang, Armour research foundation report, Tech. rep.,
  WADD-TR-60-904 (1961).

\bibitem{kats1985proprietes}
S.~A. Kats, V.~Y. Chekhovskoi, M.~D. Kovalenko, Thermophysical properties of
  zirconium and hafnium at high temperatures, Teplofizika Vysokikh Temperatur
  23~(2) (1985) 395--397.

\bibitem{qin1997spectral}
J.~Qin, M.~R{\"o}sner-Kuhn, K.~Drewes, U.~Thiedemann, G.~Kuppermann, B.~Camin,
  R.~Blume, M.~G. Frohberg, Spectral emissivities at wavelengths in the range
  500--653 nm, enthalpies, and heat capacities of the liquid phases of cobalt,
  titanium, and zirconium, High temperature and materials science 37~(3) (1997)
  129--142.

\bibitem{elyutin1967determining}
V.~Elyutin, M.~Maurakh, G.~Sverdlov, Determining the latent heat of fusion of
  zirconium., Izv. Vyssh. Ucheb. Zaved., Tsvet. Met., No. 2, 87-8 (1967).
  (1967).

\bibitem{Desai:JPCRD:1984}
P.~D. Desai, H.~M. James, C.~Y. Ho,
  \href{https://doi.org/10.1063/1.555724}{Electrical resistivity of vanadium
  and zirconium}, Journal of Physical and Chemical Reference Data 13~(4) (1984)
  1097--1130.
\newblock \href {https://doi.org/10.1063/1.555724}
  {\path{doi:10.1063/1.555724}}.

\bibitem{Peletskii:ME:1981}
V.~E. Peletskii, E.~A. Bel'skaya, Elektricheskoe soprotivlenie tugoplavkikh
  metallov (Electrical Resistance of Refractory Metals), Moscow: Energoizdat,
  1981.

\bibitem{Milosevic:IJT:2006}
N.~D. Milo{\v{s}}evi{\'{c}}, K.~D. Magli{\'{c}}, Thermophysical properties of
  solid phase zirconium at high temperatures, Int. J. Thermophys. 27~(4) (2006)
  1140--1159.
\newblock \href {https://doi.org/10.1007/s10765-006-0080-z}
  {\path{doi:10.1007/s10765-006-0080-z}}.

\bibitem{Petrova:HT:2000}
I.~I. Petrova, V.~E. Peletskii, B.~N. Samsonov, Investigation of the
  thermophysical properties of zirconium by subsecond pulsed heating technique,
  High Temp. 38~(4) (2000) 560--565.
\newblock \href {https://doi.org/10.1007/bf02755802}
  {\path{doi:10.1007/bf02755802}}.

\bibitem{Peletskii:TVT:1970}
V.~E. Peletskii, V.~Druzhinin, Y.~G. Sobol', Emissivity, thermal conductivity,
  and electrical conductivity of remelted zirconium at high temperatures,
  Teplofiz. Vys. Temp. 8~(4) (1970) 774--779.

\bibitem{Furman:1960}
S.~C. Furman, P.~A. McManus, Metal-water reactions: Ix. the kinetics of
  metal-water reactions- feasibility study of some new techniques, Tech. Rep.
  GEAP-3338, General Electric Co. Vallecitos Atomic Lab., Pleasanton, Calif.
  (1960).

\bibitem{Cubicciotti:JACS:1951}
D.~Cubicciotti, The melting point---composition diagram of the
  zirconium---oxygen system1, Journal of the American Chemical Society 73~(5)
  (1951) 2032--2035.

\bibitem{Koch::1973}
R.~K. Koch, J.~Hoffman, R.~A. Beall, Liquid State Spectral Emissivities of
  Titanium, Vanadium, Zirconium, Molybdenum, and Platinum in an Electron-beam
  Furnace, Vol. 7743, US Department of Interior, Bureau of Mines, 1973.

\bibitem{Cagran:HTP:2002}
C.~Cagran, C.~Brunner, A.~Seifter, G.~Pottlacher, Liquid-phase behaviour of
  normal spectral emissivity at 684.5~nm of some selected metals, High Temp. -
  High Pressures 34~(6) (2002) 669--679.
\newblock \href {https://doi.org/10.1068/htjr067} {\path{doi:10.1068/htjr067}}.

\bibitem{Krishnan_1994}
S.~Krishnan, C.~D. Anderson, P.~C. Nordine, Optical properties of liquid and
  solid zirconium, Phys. Rev. B 49~(5) (1994) 3161--3166.
\newblock \href {https://doi.org/10.1103/PhysRevB.49.3161}
  {\path{doi:10.1103/PhysRevB.49.3161}}.

\bibitem{Cezairliyan:IJT:1994}
A.~Cezairliyan, J.~McClure, A.~Miiller, Radiance temperatures (in the
  wavelength range 523--907 nm) of group ivb transition metals titanium,
  zirconium, and hafnium at their melting points by a pulse-heating technique,
  International journal of thermophysics 15~(5) (1994) 993--1009.

\bibitem{Coffman:P:1963}
J.~A. Coffman, G.~M. Kibler, T.~F. Lyon, B.~D. Acchione, Carbonization of
  plastics and refractory materials research, Tech. Rep. WADD-TR-60-646
  (Pt.II); AD-2979, General Electric Co. Flight Propulsion Lab. Dept.,
  Cincinnati (1961).

\bibitem{French:PRB:2014}
M.~French, T.~R. Mattsson,
  \href{https://link.aps.org/doi/10.1103/PhysRevB.90.165113}{{Thermoelectric
  transport properties of molybdenum from ab initio simulations}}, Physical
  Review B 90~(16) (2014) 165113.
\newblock \href {https://doi.org/10.1103/PhysRevB.90.165113}
  {\path{doi:10.1103/PhysRevB.90.165113}}.

\bibitem{Petrov:JL:2016}
Y.~V. Petrov, K.~P. Migdal, N.~Inogamov, S.~I. Anisimov, Transfer processes in
  a metal with hot electrons excited by a laser pulse, JETP letters 104~(6)
  (2016) 431--439.

\bibitem{Desjarlais:PRE:2017}
M.~P. Desjarlais, C.~R. Scullard, L.~X. Benedict, H.~D. Whitley, R.~Redmer,
  {Density-functional calculations of transport properties in the
  non-degenerate limit and the role of electron-electron scattering}, Physical
  Review E 95 (2017) 033203.
\newblock \href {https://doi.org/10.1103/PhysRevE.95.033203}
  {\path{doi:10.1103/PhysRevE.95.033203}}.

\bibitem{Cezairliyan1975}
A.~Cezairliyan, F.~Righini, Measurement of melting point, radiance temperature
  (at melting point), and electrical resistivity (above 2,100 k of zirconium by
  a pulse heating method, Rev. Int. Hautes Temp. Refract. 12~(3) (1975)
  201--207.

\bibitem{Ronchi_1992}
C.~Ronchi, J.~P. Hiernaut, G.~J. Hyland, Emissivity x points in solid and
  liquid refractory transition metals, Metrologia 29~(4) (1992) 261--271.
\newblock \href {https://doi.org/10.1088/0026-1394/29/4/001}
  {\path{doi:10.1088/0026-1394/29/4/001}}.

\bibitem{Baria:MT:1974}
D.~N. Baria, R.~G. Bautista, Effect of surface conditions on the normal
  spectral emittance of titanium and zirconium between 1000 and 1800 k, Metall.
  Trans. 5~(3) (1974) 555--560.
\newblock \href {https://doi.org/10.1007/bf02644649}
  {\path{doi:10.1007/bf02644649}}.

\bibitem{Cezairliyan:JRNBSSPC:1974}
A.~Cezairliyan, F.~Righini, Simultaneous measurements of heat capacity,
  electrical resistivity and hemispherical total emittance by a pulse heating
  technique: zirconium, 1500 to 2100 k, J. Res. Natl. Bur. Stand., Sect. A
  78A~(4) (1974) 509.
\newblock \href {https://doi.org/10.6028/jres.078a.034}
  {\path{doi:10.6028/jres.078a.034}}.

\bibitem{Savvatimskii:EI:2012}
A.~I. Savvatimskii, V.~N. Korobenko, High-Temperature Properties of Metals of
  Atomic Power Engineering (Zirconium, Hafnium, and Iron in Melting and Liquid
  State)[in russian], Izd. dom MEI, 2012.

\bibitem{Radousky:PRR:2020}
H.~B. Radousky, M.~R. Armstrong, R.~A. Austin, E.~Stavrou, S.~Brown, A.~A.
  Chernov, A.~E. Gleason, E.~Granados, P.~Grivickas, N.~Holtgrewe, H.~J. Lee,
  S.~S. Lobanov, B.~Nagler, I.~Nam, V.~Prakapenka, C.~Prescher, P.~Walter,
  A.~F. Goncharov, J.~L. Belof,
  \href{https://link.aps.org/doi/10.1103/PhysRevResearch.2.013192}{Melting and
  refreezing of zirconium observed using ultrafast x-ray diffraction}, Phys.
  Rev. Research 2 (2020) 013192.
\newblock \href {https://doi.org/10.1103/PhysRevResearch.2.013192}
  {\path{doi:10.1103/PhysRevResearch.2.013192}}.

\bibitem{Pigott:JPCM:2020}
J.~S. Pigott, N.~Velisavljevic, E.~K. Moss, N.~Draganic, M.~K. Jacobsen,
  Y.~Meng, R.~Hrubiak, B.~T. Sturtevant,
  \href{https://doi.org/10.1088/1361-648x/ab8cdb}{Experimental melting curve of
  zirconium metal to 37 {GPa}}, Journal of Physics: Condensed Matter 32~(35)
  (2020) 355402.
\newblock \href {https://doi.org/10.1088/1361-648x/ab8cdb}
  {\path{doi:10.1088/1361-648x/ab8cdb}}.

\bibitem{Smirnova:CMS:2018}
D.~Smirnova, S.~Starikov, I.~Gordeev, Evaluation of the structure and
  properties for the high-temperature phase of zirconium from the atomistic
  simulations, Computational Materials Science 152 (2018) 51--59.

\bibitem{Stutzmann:PRB:2015}
V.~Stutzmann, A.~Dewaele, J.~Bouchet, F.~m.~c. Bottin, M.~Mezouar,
  \href{https://link.aps.org/doi/10.1103/PhysRevB.92.224110}{High-pressure
  melting curve of titanium}, Phys. Rev. B 92 (2015) 224110.
\newblock \href {https://doi.org/10.1103/PhysRevB.92.224110}
  {\path{doi:10.1103/PhysRevB.92.224110}}.

\end{thebibliography}

%\emergencystretch=1em

\end{document}